\documentclass[11pt]{article}
\usepackage[margin=1in]{geometry}
\usepackage{amsmath,amssymb,bm,mathtools}
\usepackage{microtype}
\usepackage{physics}
\usepackage{cite}
\usepackage{graphicx}
\usepackage{url}
\usepackage{caption}
\usepackage{enumitem}
\usepackage{array}
\usepackage{xcolor}
\definecolor{QGLBlue}{RGB}{65,105,225}

\usepackage[
  unicode,
  colorlinks=true,
  pdfstartview=FitV,
  linkcolor=QGLBlue,
  citecolor=QGLBlue,
  urlcolor=QGLBlue,
  breaklinks=true
]{hyperref}

\usepackage{orcidlink}

\usepackage{graphicx}
\usepackage{url}
\usepackage{caption}
\usepackage{enumitem}
\usepackage{array}
\newcolumntype{P}[1]{>{\raggedright\arraybackslash}p{#1}}
\newcommand{\missingfigurebox}[2]{%
  \begingroup
  \setlength{\fboxsep}{0pt}%
  \fbox{%
    \begin{minipage}[c][0.23\textheight][c]{#1}
      \centering
      \vspace{0.75em}
      \textbf{Figure placeholder}\\[0.45em]
      \footnotesize Missing external file:\\
      \texttt{\detokenize{#2}}\\[0.45em]
      Standalone compilation inserts this box when the generated PNG is absent.\\
      Replace it with the exported figure for the submission build.
      \vspace{0.75em}
    \end{minipage}%
  }%
  \endgroup
}

\let\OldIncludeGraphics\includegraphics
\newcommand{\safeincludegraphics}[2][]{%
  \IfFileExists{#2}{\OldIncludeGraphics[#1]{#2}}{\missingfigurebox{0.88\linewidth}{#2}}%
}

\title{Quantum Energy Teleportation Across Lattice and Continuum}
\author{Kazuki Ikeda\,\orcidlink{0000-0003-3821-2669}\\
Department of Physics, University of Massachusetts, Boston, MA 02125, USA}
\date{}

\begin{document}
\maketitle

\begin{abstract}
Quantum energy teleportation (QET) has been studied in continuum field theory and in lattice many-body systems, but the relation between the two within a single interacting model is still not well understood.  To address this question, we consider the massive Thirring model, equivalently the sine--Gordon theory.  In the continuum, the trigonometric measurement is a weak binary Positive Operator-Valued Measure (POVM), and its leading signal is set by a conserved-current correlator in the bosonized theory, with both gapless behavior and gapped large-distance asymptotics.  On the lattice, the conventional protocol does not access this neutral current sector. For Alice's local measurement, a lattice $U(1)$ selection rule removes the neutral current contribution from Bob's subsystem, and the separated signal lies in charged sectors.  On the same lattice Hamiltonian we construct a neutral current protocol whose weak signal is exactly a coarse-grained current correlator and whose extracted energy scales quadratically with the measurement strength.  This identifies the neutral sector shared by the lattice and continuum descriptions, while separating it from the charged sector that governs the conventional qubit protocol.
\end{abstract}

\setcounter{tocdepth}{2} 
\tableofcontents\

\section{Introduction}
Quantum energy teleportation (QET) uses quantum correlations already present in a ground/vacuum state to extract energy at a distant site through a local measurement, classical communication, and feed-forward operations \cite{Hotta2008,HottaSpinChain2009}. (Detailed reviews are available in.~\cite{Hotta:2011xj,IkedaQuantumEnergyComputers2025,RagulaMartinMartinez2025}.) Experimental demonstrations were recently conducted at finite temperature \cite{RodriguezBrionesEtAl2023} and near zero temperature \cite{IkedaHardware2023}.  The subject spans relativistic field theory and lattice many-body systems.  In relativistic field theory, local algebraic structure, passivity, and constraints on negative energy provide the basic framework \cite{SummersWernerI1987,SummersWernerII1987,PuszWoronowicz1978,Ford1991,FewsterEveson1998}.  In many-body physics, the same protocol is tied to critical correlations and entanglement structure \cite{CalabreseCardy2004,EisertCramerPlenio2010,CasiniHuerta2009}.  Finite-dimensional realizations and operational bounds were examined in \cite{AlhambraEtAl2019}. 

Continuum and near-continuum protocols were developed for trapped ions, electromagnetic fields, harmonic chains, and quantum Hall edge channels \cite{HottaTrappedIons2009,HottaEM2010,NambuHotta2010,YusaIzumidaHotta2011}.  The relation between teleported energy and entanglement resources, including long distance variants based on squeezed states, was studied in \cite{HottaEnergyEntanglement2010,HottaEntanglementBound2013,HottaDistance2014}.  Extensions to relativistic protocols with qudit probes, negative stress-energy generated through QET, and quasi-local measurements in two-dimensional conformal field theory were analyzed in \cite{VerdonAkzamMartinMartinezKempf2016,FunaiMartinMartinez2017,GuoLin2018}.  Recent work has also clarified local limits on extraction, strong local passive states, experimental activation, information-theoretic formulations of minimal protocols, and upper bounds on extractable energy \cite{AlhambraEtAl2019,WuEtAl2024,FanEtAlStrong2024,FanEtAlResources2024,WangYao2024,ikeda2025quantum,IkedaBeyond2025,2mkn-rjff,Haque:2026pxz}.

QET has also become a probe of many-body structure and control.  Examples include criticality in the massive Thirring model \cite{Ikeda2023}, symmetry-protected topological phase diagrams \cite{IkedaSPT2023}, robustness of post-measurement quantum correlations \cite{IkedaLoweRobustness2024}, quantum networks \cite{IkedaHyperbolic2024}, quantum cryptographic settings \cite{IkedaQIP2024,n7t7-2zvw,7s7y-7xcn}, quantum batteries \cite{HottaIkedaBattery2025}, game theory \cite{ikeda2025quantum}, and feedback activation of observables beyond energy, including current and charge \cite{IkedaBeyond2025}.

Yet continuum and lattice QET are usually written in different operator languages.  The continuum construction is naturally formulated with local POVMs, while lattice studies often use projective qubit measurements.  In an interacting theory this is not a superficial difference.  After bosonization and open boundaries are imposed, it is not obvious which lattice sector should reproduce a given continuum observable.

The massive Thirring model is a natural place to examine this question.  Through bosonization it is equivalent to the sine--Gordon theory, and the conserved fermion current becomes a local bosonic field \cite{Coleman1975,Mandelstam1975}.  In the continuum construction of Refs.~\cite{Hotta2008,HottaEM2010,NambuHotta2010}, Alice's measurement is a weak binary POVM in the standard quantum-information sense: for small measurement strength, each Kraus operator stays close to the identity and the leading QET coefficient is set by a smeared current correlator.  This is not a weak-value protocol with an auxiliary meter and postselection.

On the lattice, the open-chain protocol of Ref.~\cite{Ikeda2023} uses single-site Pauli measurements and local Pauli feedback on the same massive Thirring Hamiltonian.  Those measurements are projective.  The small parameter there is the weak signal ratio that governs the optimized extracted energy.  In the original Alice-$X_{n_A}$ circuit, a lattice $U(1)$ selection rule removes the Bob-side neutral current channel, so the separated signal lies in charged sectors.  Related chirality-based projective protocols can activate local current and charge on the lattice \cite{IkedaBeyond2025}, but the protocol studied here isolates the neutral conserved current channel and matches it to the continuum weak binary POVM.

A second protocol on the same lattice Hamiltonian, built from coarse-grained current observables, restores the neutral channel and yields a weak signal that is exactly the corresponding lattice current correlator.  This gives a direct neutral-sector comparison between lattice and continuum, while the charged part of the original qubit protocol is treated separately and checked against a boundary sign test.

The rest of this article is organized as follows.  Section~\ref{sec:continuumHotta} recalls the continuum QET protocol in canonical normalization and fixes the weak binary POVM together with its weak-measurement expansion.  Section~\ref{sec:bosonizedCurrent} rewrites the leading continuum coefficient in bosonized massive Thirring variables and gives its exact bulk form-factor representation.  Sections~\ref{sec:gaplessContinuum} and~\ref{sec:gappedBulkSG} then treat the continuum theory in the gapless and gapped regimes, including the open-boundary gapless kernel and the large-distance massive asymptotics.  Section~\ref{sec:latticeConnection} turns to the open-chain lattice problem. It first analyzes the conventional QET protocol, then introduces a modified neutral-current protocol on the same Hamiltonian, and finally compares the remaining charged sector with a boundary description. Section~\ref{sec:discussion} discusses the sector structure of the comparison, Section~\ref{sec:conclusions} summarizes the main results, and Appendix~\ref{app:LZnormalization} records the relation to the Lukyanov--Zamolodchikov normalization.

\section{Continuum QET protocol in canonical normalization}
\label{sec:continuumHotta}

\subsection{Canonical field, local energy, and Bob's local unitary}
We use a canonically normalized bosonic field $\phi$ and its conjugate momentum $\Pi$:
\begin{equation}
[\phi(t,x),\Pi(t,y)]=i\,\delta(x-y).
\end{equation}
Throughout the continuum field-theory analysis we work in relativistic units with propagation speed set to $1$, so every occurrence of $T$ below is the relativistic continuum time coordinate.

For the bosonized SG theory we take the Hamiltonian density in canonical form,
\begin{equation}
\mathcal H(x)=\frac12\Pi(x)^2+\frac12(\partial_x\phi(x))^2+2\mu\bigl(1-\cos(\beta_C\phi(x))\bigr),
\label{eq:HcanSG}
\end{equation}
with the additive constant chosen so that the ground-state energy density vanishes.  Let $w_B(x)$ be a nonnegative window function supported in Bob's region and define
\begin{equation}
\hat H_B\equiv \int dx\,w_B(x)\,\hat\varepsilon(x),
\end{equation}
where $\hat\varepsilon$ is the normal-ordered local energy density associated with \eqref{eq:HcanSG}.  We assume $w_B(x)=1$ on the support of Bob's smearing profile $p_B(x)$.

Bob's outcome-dependent local unitary is the local shift operator used in Ref.~\cite{Hotta2008},
\begin{equation}
\hat U_n(B)=\exp\!\biggl(i\,\kappa a_n\int dx\,p_B(x)\,\hat\phi(x)\biggr),
\label{eq:BobUnitary}
\end{equation}
where $a_n\in\mathbb R$ labels the classical outcome and $\kappa$ is Bob's control parameter.  By the Baker--Campbell--Hausdorff formula,
\begin{equation}
\hat U_n(B)^\dagger\,\hat\Pi(x)\,\hat U_n(B)=\hat\Pi(x)+\kappa a_n p_B(x).
\end{equation}
Therefore
\begin{equation}
\hat U_n(B)^\dagger\hat H_B\hat U_n(B)=\hat H_B+\kappa a_n\hat O_B+\frac{\kappa^2 a_n^2}{2}\int dx\,w_B(x)p_B(x)^2,
\label{eq:HBshift}
\end{equation}
where
\begin{equation}
\hat O_B\equiv \int dx\,w_B(x)p_B(x)\,\hat\Pi(x).
\label{eq:OBdef}
\end{equation}
The factor $1/2$ in the quadratic term is the one appropriate to the canonical normalization \eqref{eq:HcanSG}.

\subsection{Trigonometric POVM}
Alice uses a two-valued POVM localized in region $A$:
\begin{align}
\hat M_0(A)&=\cos\hat\Phi_A,
&
\hat M_1(A)&=\sin\hat\Phi_A,
\label{eq:HottaPOVM}
\end{align}
with
\begin{equation}
\hat\Phi_A\equiv \frac{\pi}{4}-\hat X_A,
\qquad
\hat X_A\equiv \lambda_0\int dx\,f_A(x)\,\hat\Pi(0,x),
\label{eq:PhiAndX}
\end{equation}
where $f_A$ is supported in $A$ and $\lambda_0$ controls the measurement strength.
Define the POVM elements $\hat D_n=\hat M_n^\dagger\hat M_n$ and outcome labels $a_n=(-1)^n$.  Then
\begin{align}
\hat D_A &\equiv \sum_n a_n\hat D_n,
\label{eq:DAdef}
\\
\hat{\tilde D}_A^{\,2}&\equiv \sum_n a_n^2\hat D_n.
\label{eq:Dtildedef}
\end{align}
For \eqref{eq:HottaPOVM} one has the exact identities
\begin{equation}
\hat D_A=\hat D_0-\hat D_1=\cos(2\hat\Phi_A)=\sin(2\hat X_A),
\qquad
\hat{\tilde D}_A^{\,2}=\hat D_0+\hat D_1=\mathbf 1.
\label{eq:DAtildeExact}
\end{equation}
The coefficient multiplying $\kappa^2$ depends on $\hat{\tilde D}_A^{\,2}$ rather than $\hat D_A^2$, and for the present POVM one has $\hat{\tilde D}_A^{\,2}=\mathbf 1$ exactly.

\subsection{Weak binary POVM}
The measurement in Eqs.~\eqref{eq:HottaPOVM} and \eqref{eq:PhiAndX} is weak in the standard quantum-information sense of an unsharp two-outcome POVM.  Expanding for small $\lambda_0$ gives
\begin{equation}
\hat M_0(A)=\frac{1}{\sqrt 2}\bigl(\mathbf 1+\hat X_A\bigr)+O(\lambda_0^2),
\qquad
\hat M_1(A)=\frac{1}{\sqrt 2}\bigl(\mathbf 1-\hat X_A\bigr)+O(\lambda_0^2).
\label{eq:weakMexpansion}
\end{equation}
The associated POVM elements therefore obey
\begin{equation}
\hat D_0=\frac12\mathbf 1+\hat X_A+O(\lambda_0^3),
\qquad
\hat D_1=\frac12\mathbf 1-\hat X_A+O(\lambda_0^3).
\label{eq:weakDexpansion}
\end{equation}
Each Kraus operator is thus close to the identity and the outcome bias is linear in $\hat X_A$.  This is the usual weak or unsharp binary measurement.

This is not a weak-value scheme with a separate meter and postselection.  No auxiliary pointer system enters, and there is no postselection.  The single-site Pauli measurement of Ref.~\cite{Ikeda2023} is projective rather than weak in the POVM sense.  On the lattice side, the small parameter appears only after Bob's optimization through the signal ratio $|\eta_{\rm lat}|/\xi_{\rm lat}\ll1$ in Eq.~\eqref{eq:EBlatWeak}.  The neutral current lattice protocol of Sec.~\ref{sec:latticeConnection} is the direct lattice analogue of the continuum weak binary POVM.

\subsection{Spacelike timing and the quadratic form}
Let $\hat U(T)=e^{-i\hat H T}$ be the system time evolution.  We impose a spacelike protocol,
\begin{equation}
 T < L-(R_A+R_B),
\label{eq:spacelike_cond}
\end{equation}
where $L$ is the separation between the supports of $A$ and $B$ and $R_{A,B}$ denote their radii.  Then every operator supported in $A$ at $t=0$ commutes with every operator supported in $B$ at $t=T$.  In particular, the commutator conditions needed for the quadratic form follow directly from microcausality in the spacelike regime \cite{Hotta2008}.  In this formulation the classical communication step is treated as an external side channel, while the field theory governs the measurement and feedback operators themselves.

Averaging over outcomes after Bob's conditional unitary, one obtains
\begin{equation}
\expval{\hat H_B}_{\rho_F}=\frac12\,\xi\,\kappa^2+\eta\,\kappa,
\label{eq:HottaQuadratic}
\end{equation}
with
\begin{align}
\xi&\equiv \expval{\hat{\tilde D}_A^{\,2}}_g\int dx\,w_B(x)p_B(x)^2,
\label{eq:xi_def}
\\
\eta&\equiv \expval{\hat D_A\,\hat O_B(T)}_g,
\qquad
\hat O_B(T)\equiv \hat U(T)^\dagger\hat O_B\hat U(T).
\label{eq:eta_def}
\end{align}
Because \eqref{eq:DAtildeExact} gives $\expval{\hat{\tilde D}_A^{\,2}}_g=1$, we have
\begin{equation}
\xi=\int dx\,w_B(x)p_B(x)^2.
\label{eq:xi_simple}
\end{equation}
Optimizing \eqref{eq:HottaQuadratic} over $\kappa$ yields
\begin{equation}
\kappa_*=-\frac{\eta}{\xi},
\qquad
E_B^{\max}=\frac{\eta^2}{2\xi}.
\label{eq:EBmax}
\end{equation}

\subsection{Weak-measurement expansion}
For the interacting SG vacuum, the remaining step beyond the exact quadratic form is the weak measurement expansion,
\begin{equation}
\hat D_A=\sin(2\hat X_A)=2\hat X_A+O(\lambda_0^3).
\label{eq:weakDA}
\end{equation}
Hence
\begin{equation}
\eta=2\expval{\hat X_A\hat O_B(T)}_g+O(\lambda_0^3).
\end{equation}
Substituting \eqref{eq:PhiAndX} and \eqref{eq:OBdef} gives the exact leading-order formula
\begin{equation}
\eta
=2\lambda_0\int dx\,dy\,f_A(x)w_B(y)p_B(y)
\expval{\Pi(0,x)\Pi(T,y)}_g
+O(\lambda_0^3).
\label{eq:etaPiPi}
\end{equation}
Writing
\begin{equation}
\eta(\lambda_0)=\eta_1\lambda_0+O(\lambda_0^3),
\qquad
\eta_1\equiv 2\int dx\,dy\,f_A(x)w_B(y)p_B(y)\expval{\Pi(0,x)\Pi(T,y)}_g,
\label{eq:eta1def}
\end{equation}
one obtains the leading extracted energy
\begin{equation}
E_B^{\max}(\lambda_0)=\frac{\eta_1^2}{2\xi}\,\lambda_0^2+O(\lambda_0^4).
\label{eq:EB_lambda_expansion}
\end{equation}
This expresses the leading QET coefficient entirely in terms of the interacting current correlator.  No Gaussian identities are used inside the interacting SG vacuum.

\section{Bosonization of the massive Thirring current sector}
\label{sec:bosonizedCurrent}
\subsection{Canonical SG action, Coleman relation, and current map}
The bosonized massive Thirring theory \cite{Thirring:1958in}, equivalently the sine--Gordon theory, provides a natural setting in which to compare continuum and lattice descriptions. See textbook \cite{Korepin_Bogoliubov_Izergin_1993} for the details about the model.

We work throughout with the canonically normalized SG field,
\begin{equation}
S_{\rm SG}=\int d^2x\,\Bigl[\frac12(\partial_\mu\phi)^2-2\mu\bigl(1-\cos(\beta_C\phi)\bigr)\Bigr].
\label{eq:SGcanonicalAction}
\end{equation}
The bosonized Thirring current is
\begin{equation}
 j^\mu = -\frac{\beta_C}{2\pi}\,\epsilon^{\mu\nu}\partial_\nu\phi,
\label{eq:currentmapCanonical}
\end{equation}
so that
\begin{equation}
 j^1=\frac{\beta_C}{2\pi}\,\Pi,
\qquad
\Pi=\frac{2\pi}{\beta_C}\,j^1.
\label{eq:PiCurrentCanonical}
\end{equation}
The Coleman relation takes the standard form
\begin{equation}
\frac{4\pi}{\beta_C^2}=1+\frac{g}{\pi}.
\label{eq:ColemanCanonical}
\end{equation}
The cosine perturbation has scaling dimension $\Delta_{\cos}=\beta_C^2/(4\pi)$ and is therefore marginal at
\begin{equation}
\beta_C^2=8\pi
\qquad\Longleftrightarrow\qquad
 g=-\frac{\pi}{2}.
\label{eq:BKTcanonical}
\end{equation}
This is the BKT line in Coleman's canonical normalization.

Via \eqref{eq:PiCurrentCanonical}, the leading QET coefficient becomes a functional of the current correlator,
\begin{equation}
\eta_1
=\frac{8\pi^2}{\beta_C^2}\int dx\,dy\,f_A(x)w_B(y)p_B(y)
\expval{j^1(0,x)j^1(T,y)}_g.
\label{eq:eta1current}
\end{equation}
The weak measurement signal is therefore controlled by a bosonized Thirring current correlator.

\subsection{Implementation in bosonized Thirring variables}
At leading order in the weak measurement, the continuum protocol can be implemented directly in bosonized Thirring variables without further reference to lattice observables.  Writing
\begin{equation}
\hat X_A=\lambda_0\frac{2\pi}{\beta_C}\int dx\,f_A(x)\,\hat j^1(0,x),
\qquad
\hat O_B(T)=\frac{2\pi}{\beta_C}\int dy\,g_B(y)\,\hat j^1(T,y),
\label{eq:currentSectorRecipe}
\end{equation}
with $g_B(y)\equiv w_B(y)p_B(y)$, the continuum construction in the current sector may be summarized as follows.
\begin{enumerate}[label=(\alph*)]
\item Choose the SG and Thirring couplings $(\beta_C,\mu)$ and smooth compactly supported smearings $f_A,g_B$ with spacelike-separated supports.
\item Evaluate the exact leading-order coefficient $\eta_1(L,T)$ either from the current correlator \eqref{eq:eta1current} or from the bulk spectral formula \eqref{eq:eta1ExactFF} below.
\item Compute $\xi=\int dx\,w_B(x)p_B(x)^2$ and then
\begin{equation}
\kappa_*= -\frac{\eta_1}{\xi}\lambda_0+O(\lambda_0^3),
\qquad
E_B^{\max}(\lambda_0)=\frac{\eta_1^2}{2\xi}\lambda_0^2+O(\lambda_0^4).
\label{eq:kappaStarRecipe}
\end{equation}
\end{enumerate}
Within the continuum protocol and the standard spectral and form-factor formalism for the current sector, the bosonized massive Thirring implementation is exact at the level of the leading weak measurement coefficient $\eta_1$ and its representation \eqref{eq:eta1ExactFF}.  The protocol is already written in bosonized Thirring current variables through \eqref{eq:currentSectorRecipe}, while the SG field provides a convenient representation of the same current sector Wightman function.  In the gapped theory, only the further reduction to large distance asymptotics is approximate.

\subsection{\texorpdfstring{Exact bulk form-factor implementation of $\eta_1$}{Exact bulk form-factor implementation of eta1}}
Eq. \eqref{eq:eta1current} is exact at leading order in $\lambda_0$ for any SG vacuum.  In the bulk gapped theory, translation invariance allows a direct form-factor implementation.  Let Alice's smearing be centered at the origin and write Bob's local weight as
\begin{equation}
g_B(y)\equiv w_B(y)p_B(y)=g(y-L),
\end{equation}
with $L>0$ the center-to-center separation.  Define the Fourier transforms
\begin{equation}
\tilde f(q)=\int dx\,e^{-iqx}f_A(x),
\qquad
\tilde g(q)=\int du\,e^{-iqu}g(u).
\label{eq:FTdefs}
\end{equation}
The bulk Wightman function of the current has the exact spectral expansion
\begin{equation}
\expval{j^1(0,0)j^1(T,z)}_g
=\sum_{n=1}^{\infty}\frac{1}{n!}
\sum_{a_1,\dots,a_n}
\int\frac{d\theta_1\cdots d\theta_n}{(2\pi)^n}
\left|F_n^{j^1\,|\,a_1\dots a_n}(\theta_1,\dots,\theta_n)\right|^2
 e^{-iT E_n(\bm\theta)+iz P_n(\bm\theta)},
\label{eq:currentFFWightman}
\end{equation}
where
\begin{equation}
E_n(\bm\theta)=\sum_{k=1}^{n}m_{a_k}\cosh\theta_k,
\qquad
P_n(\bm\theta)=\sum_{k=1}^{n}m_{a_k}\sinh\theta_k.
\end{equation}
Substituting \eqref{eq:currentFFWightman} into \eqref{eq:eta1current} and performing the $x,y$ integrals yields the exact leading-order bulk implementation formula
\begin{equation}
\eta_1(L,T)
=\frac{8\pi^2}{\beta_C^2}
\sum_{n=1}^{\infty}\frac{1}{n!}
\sum_{a_1,\dots,a_n}
\int\frac{d\theta_1\cdots d\theta_n}{(2\pi)^n}
\left|F_n^{j^1\,|\,a_1\dots a_n}(\bm\theta)\right|^2
 e^{-iT E_n(\bm\theta)+iL P_n(\bm\theta)}
 \tilde f\!\bigl(P_n(\bm\theta)\bigr)
 \tilde g\!\bigl(-P_n(\bm\theta)\bigr).
\label{eq:eta1ExactFF}
\end{equation}
Equation \eqref{eq:eta1ExactFF} gives the exact continuum QET formula in the current sector of the bosonized Thirring theory at leading order in the measurement strength.  All later gapped asymptotics are derived from this expression.  For smooth compactly supported smearings, $\tilde f$ and $\tilde g$ are entire and rapidly decaying, so the spectral integrals are well behaved.

\section{Gapless continuum implementation}
\label{sec:gaplessContinuum}
\subsection{Exact full-line kernel and fixed rapidity law}
At the gapless fixed point, the canonical momentum correlator on the full line is
\begin{equation}
\langle \Pi(0,x)\Pi(T,y)\rangle_0
=-\frac{1}{4\pi}\left[\frac{1}{(x-y-T)^2}+\frac{1}{(x-y+T)^2}\right],
\label{eq:masslessPiPi}
\end{equation}
valid in the spacelike regime $|x-y|>|T|$ as a distribution away from coincident points.  Substituting \eqref{eq:masslessPiPi} into \eqref{eq:etaPiPi} shows that, for smearings localized on scales much smaller than the separation,
\begin{equation}
\eta_1(L,T)\propto \frac{1}{(L-T)^2}+\frac{1}{(L+T)^2}
=\frac{2(L^2+T^2)}{(L^2-T^2)^2}
=\frac{2\cosh(2\chi)}{r^2},
\label{eq:etaGaplessScaling}
\end{equation}
where we parameterize the spacelike separation as
\begin{equation}
L=r\cosh\chi,
\qquad
T=r\sinh\chi,
\qquad
r\equiv \sqrt{L^2-T^2}>0.
\label{eq:spacelikeRapidity}
\end{equation}
Hence
\begin{equation}
E_B^{\max}(L,T)\propto r^{-4}
\qquad\text{at fixed spacelike rapidity }\chi.
\label{eq:EBgaplessScaling}
\end{equation}
The hyperbolic factor $\cosh(2\chi)$ makes the approach to the light cone anisotropic even though the overall radial decay is $r^{-2}$ at fixed $\chi$.

\subsection{Compact-support continuum implementation}
\begin{figure}[t]
  \centering
  \safeincludegraphics[width=0.95\linewidth]{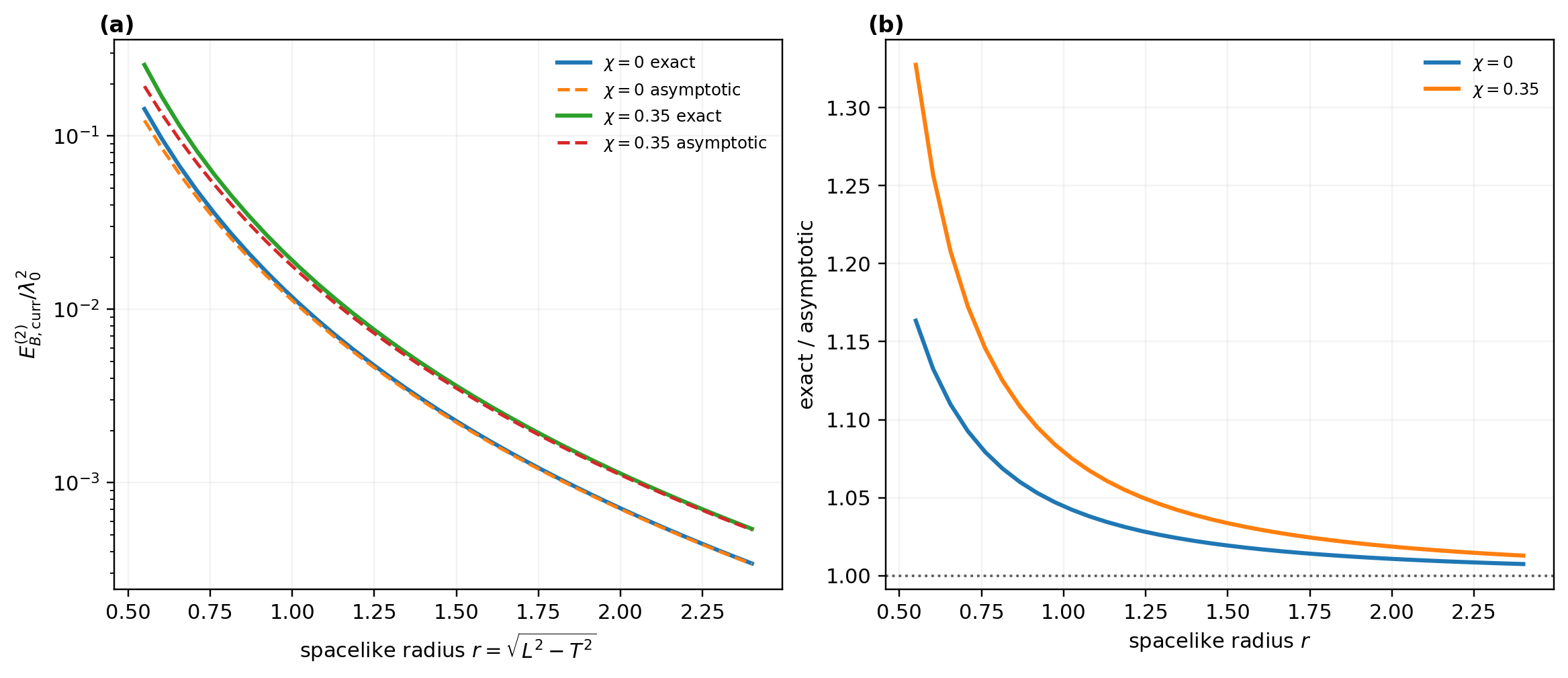}
  \caption{Direct continuum implementation of the leading weak measurement energy signal $E_{B,\rm curr}^{(2)}=\eta_1^2\lambda_0^2/2$ at the massless fixed point using the compactly supported smearings \eqref{eq:compactBumps} with $R_A=R_B=0.15$ and $\xi=1$.  (a) exact numerical evaluation of Eq.~\eqref{eq:eta1_compact} for two fixed spacelike rapidities, as well as the asymptotic form implied by Eq.~\eqref{eq:eta1_asym_compact}.  (b) ratio of the exact result to the asymptotic expression, showing convergence to the predicted fixed rapidity $r^{-4}$ tail.}
  \label{fig:gaplessImplementation}
\end{figure}

To make the continuum QET construction explicit, we choose smooth compactly supported bump profiles
\begin{equation}
\begin{aligned}
f_A(x)&=\mathcal N_A\exp\!\left[-\frac{1}{1-(x/R_A)^2}\right]\Theta(R_A-|x|),\\
g_B(y)&=\mathcal N_B\exp\!\left[-\frac{1}{1-((y-L)/R_B)^2}\right]\Theta(R_B-|y-L|),
\end{aligned}
\label{eq:compactBumps}
\end{equation}
with $\int dx\,f_A(x)=1$ and $\int dy\,g_B(y)^2=1$, so that $\xi=1$ exactly.  In the massless fixed-point theory the leading weak measurement coefficient is therefore
\begin{equation}
\eta_1(L,T)=-\frac{1}{2\pi}\int_A dx\int_B dy\,f_A(x)g_B(y)
\left[\frac{1}{(x-y-T)^2}+\frac{1}{(x-y+T)^2}\right],
\label{eq:eta1_compact}
\end{equation}
and the leading energy signal is
\begin{equation}
E_{B,\rm curr}^{(2)}(L,T)
\equiv \frac{\eta_1(L,T)^2}{2}\,\lambda_0^2.
\label{eq:EB2_compact}
\end{equation}
At large fixed rapidity one finds
\begin{equation}
\eta_1(L,T)= -\frac{P_B}{2\pi}\left[\frac{1}{(L-T)^2}+\frac{1}{(L+T)^2}\right]+O(r^{-3}),
\qquad
P_B\equiv \int dy\,g_B(y),
\label{eq:eta1_asym_compact}
\end{equation}
which reproduces \eqref{eq:EBgaplessScaling}.  Fig.~\ref{fig:gaplessImplementation} shows a direct numerical implementation of \eqref{eq:eta1_compact} and \eqref{eq:EB2_compact} with compactly supported smearings.  It provides a concrete continuum realization of the bosonized QET protocol in the Thirring current sector, independent of any lattice construction.

\subsection{Open boundaries in the gapless regime}
On a half-line, the method of images gives
\begin{equation}
G_{\Pi\Pi}^{\rm OBC}(x,y;T)
=G_{\Pi\Pi}^{\infty}(x-y;T)
+s_{\rm bc}\,G_{\Pi\Pi}^{\infty}(x+y;T),
\label{eq:OBCgaplessHalf}
\end{equation}
where $s_{\rm bc}=+1$ for Neumann and $s_{\rm bc}=-1$ for Dirichlet boundary conditions for the boson field.  On a finite interval $(0,\ell)$ one obtains the corresponding image sum over $y\to \pm y+2n\ell$.  Consequently,
\begin{equation}
\eta_{1,\rm OBC}(L,T)
\propto \frac{1}{(L-T)^2}+\frac{1}{(L+T)^2}
+s_{\rm bc}\left[\frac{1}{(L'-T)^2}+\frac{1}{(L'+T)^2}\right]+\cdots,
\label{eq:etaOBCgapless}
\end{equation}
where $L'=x_A+x_B$ is the distance to the image point and the omitted terms come from further images on a finite interval.  These formulas are derived only in the gapless theory.  Later, when the open-chain lattice geometry of Ref.~\cite{Ikeda2023} is compared with the continuum description, the exact no-flux lattice continuity equation singles out the Dirichlet choice $s_{\rm bc}=-1$ as the natural boundary condition for the neutral current sector.  For the interacting gapped theory, however, the corresponding boundary SG correlator is not yet available, so the finite-interval image kernel used below serves as a reference kernel for that comparison.

\section{Gapped bulk SG: spectral formula and large distance asymptotics}
\label{sec:gappedBulkSG}
\subsection{General bulk asymptotics from the current spectral formula}
For $\beta_C^2<8\pi$, the cosine is relevant and the SG theory is massive.  In the spacelike regime it is convenient to use the Euclidean spectral representation and then continue back to
\begin{equation}
r\equiv\sqrt{L^2-T^2}>0.
\label{eq:rdef}
\end{equation}
Let $M_{\rm eff}$ denote the smallest total mass that appears with a nonvanishing current form factor in \eqref{eq:eta1ExactFF}.  Then the exact form-factor implementation implies the large distance bulk law
\begin{equation}
\eta_1(L,T)\sim \mathcal A_j\,r^{-\gamma_j}e^{-M_{\rm eff}r},
\qquad
E_B^{\max}(L,T)\sim r^{-2\gamma_j}e^{-2M_{\rm eff}r},
\label{eq:etaEB_gapped_general}
\end{equation}
where $\gamma_j$ is the threshold exponent of the lightest current channel and $\mathcal A_j$ is a nonuniversal amplitude determined by the corresponding form factor and smearing overlap.  The leading asymptotic assumes that the chosen smearings have nonzero overlap with the lightest current channel.  If the threshold overlap vanishes by symmetry or moment constraints, then either the same channel reappears with a higher threshold power or the next allowed channel controls the right tail.

Writing
\begin{equation}
s\equiv \frac{r}{\xi_{\rm corr}^{(j)}}=M_{\rm eff}r,
\qquad
\xi_{\rm corr}^{(j)}\equiv \frac{1}{M_{\rm eff}},
\label{eq:sdef}
\end{equation}
the asymptotic regime of \eqref{eq:etaEB_gapped_general} is $s\gg 1$.  At fixed finite $r$, approaching the BKT line drives $\xi_{\rm corr}^{(j)}\to\infty$, so eventually $s\lesssim 1$ and the gapped expression crosses over to the gapless fixed-point law.  The gapped formula therefore explains the enhancement of the teleported energy on the gapped side as the current-channel correlation length increases, but not the full peak profile.

\subsection{One-particle neutral current channel}
If the lightest nonvanishing current channel is a single neutral mode of mass $m_*$, then the channel-truncated correlator can be written exactly in terms of the Bessel kernel.  Equivalently, one may think of this as an effective current mode with
\begin{equation}
\expval{j^1(x)j^1(0)}^{(1)} = Z_j\,\mathcal G_{j^1j^1}^{(1)}(x),
\end{equation}
where $Z_j$ is the channel amplitude and
\begin{equation}
\mathcal G_{j^1j^1}^{(1)}(L,T)
= -\frac{1}{2\pi}\,\partial_T^2 K_0(m_* r)
= -\frac{1}{2\pi}\left[\frac{m_*\cosh(2\chi)}{r}K_1(m_* r)+m_*^2\sinh^2\chi\,K_0(m_* r)\right]
\label{eq:oneParticleKernel}
\end{equation}
with $L=r\cosh\chi$ and $T=r\sinh\chi$.  The corresponding channel-truncated QET coefficient is
\begin{equation}
\eta_1^{(1)}(L,T)
=\frac{8\pi^2 Z_j}{\beta_C^2}\int dx\,dy\,f_A(x)g_B(y)
\mathcal G_{j^1j^1}^{(1)}(y-x,T).
\label{eq:eta1_oneparticle}
\end{equation}
It is convenient to factor out the channel amplitude as
\begin{equation}
\eta_1^{(1)}(L,T)=Z_j^{\rm eff}\,\widehat\eta_1^{(1)}(L,T),
\qquad
Z_j^{\rm eff}\equiv \frac{8\pi^2 Z_j}{\beta_C^2},
\label{eq:ZjeffDef}
\end{equation}
so that the leading weak measurement energy in the one-particle truncation is
\begin{equation}
E_{B,\rm curr}^{(2,1)}(L,T)=\frac{1}{2}\left[\eta_1^{(1)}(L,T)\right]^2\lambda_0^2
=\frac{1}{2}(Z_j^{\rm eff})^2\left[\widehat\eta_1^{(1)}(L,T)\right]^2\lambda_0^2.
\label{eq:EB_oneparticle}
\end{equation}
This statement is exact within the one-particle channel truncation of \eqref{eq:eta1ExactFF}.  In the numerical illustration below we use the equivalent $\Pi\Pi$ normalization and plot the dimensionless combination $E_{B,\rm curr}^{(2,1)}/[(Z_j^{\rm eff})^2\lambda_0^2]$, setting $Z_j^{\rm eff}=1$ only after this normalization.

Equation \eqref{eq:oneParticleKernel} immediately shows the directional asymptotics:
\begin{equation}
\mathcal G_{j^1j^1}^{(1)}(L,T)
\sim e^{-m_* r}\Bigl[A_1(\chi)\,r^{-1/2}+B_1(\chi)\,r^{-3/2}+\cdots\Bigr],
\label{eq:oneparticleDirectional}
\end{equation}
with $A_1(\chi)\propto\sinh^2\chi$ and $B_1(\chi)\propto\cosh(2\chi)$.  Therefore,
\begin{equation}
\gamma_j=\frac12
\qquad\text{for a generic fixed spacelike rapidity }\chi\neq 0,
\label{eq:gamma_generic_oneparticle}
\end{equation}
whereas on the equal-time slice $T=0$ one has $A_1(0)=0$ and therefore
\begin{equation}
\gamma_j=\frac32
\qquad\text{for the equal-time one-particle channel.}
\label{eq:gamma_equaltime_oneparticle}
\end{equation}
The limit $\chi\to0$ is nonuniform: because $A_1(\chi)\propto \sinh^2\chi$, the crossover to the generic $r^{-1/2}e^{-m_*r}$ law for small but nonzero $\chi$ occurs only at parametrically large $r$, schematically $m_*r\gtrsim O(\sinh^{-2}\chi)$.

For compactly supported smearings with support radii held fixed while the center-to-center separation $L$ becomes large, the channel-truncated compact-support integral admits the reduced approximation
\begin{equation}
\eta_{1,\rm c2c}^{(1)}(L,T)
=\frac{8\pi^2 Z_j}{\beta_C^2}\,\tilde f(0)\,\tilde g(0)\,\mathcal G_{j^1j^1}^{(1)}(L,T),
\label{eq:eta1_c2c}
\end{equation}
with the corresponding leading energy signal
\begin{equation}
E_{B,\rm curr,c2c}^{(2,1)}(L,T)=\frac12\left[\eta_{1,\rm c2c}^{(1)}(L,T)\right]^2\lambda_0^2.
\label{eq:EB_oneparticle_c2c}
\end{equation}
The pure large distance tail is the further reduction of \eqref{eq:EB_oneparticle_c2c} to the asymptotic powers listed in Eqs.~\eqref{eq:gamma_generic_oneparticle}--\eqref{eq:gamma_equaltime_oneparticle}.

\begin{figure}[t]
  \centering
  \safeincludegraphics[width=0.95\linewidth]{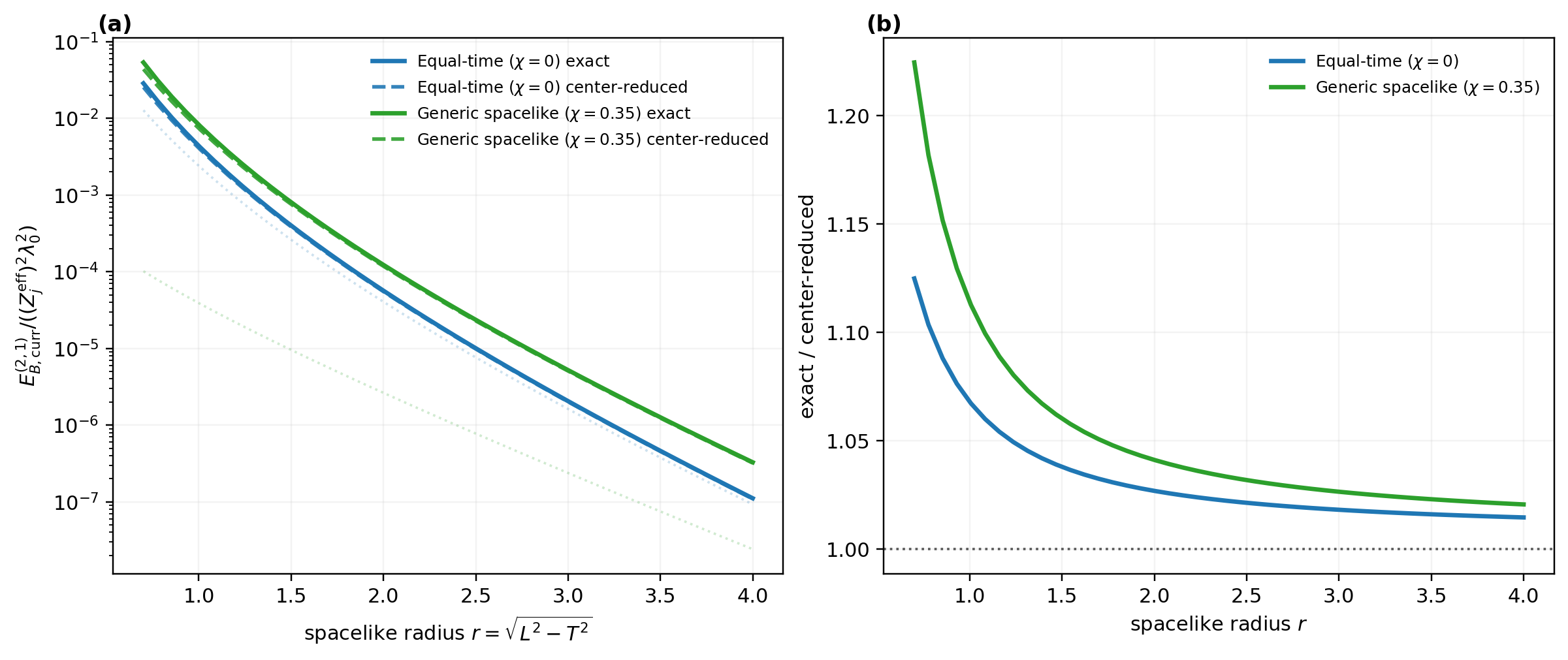}
  \caption{Compact-support implementation of the one-particle contribution to the continuum signal, using the kernel \eqref{eq:oneParticleKernel}.  (a) Numerical evaluation of \eqref{eq:eta1_oneparticle} for two fixed spacelike rapidities (solid) as well as the large distance center-to-center approximation \eqref{eq:eta1_c2c}--\eqref{eq:EB_oneparticle_c2c} (dashed).  Faint dotted curves indicate the asymptotic tail $r^{-2\gamma_j}e^{-2m_*r}$.  The plotted vertical scale is the dimensionless quantity $E_{B,\rm curr}^{(2,1)}/[(Z_j^{\rm eff})^2\lambda_0^2]$.  (b) Ratio of the exact one-particle result to the center-to-center approximation.}
  \label{fig:gappedOneParticleImplementation}
\end{figure}

To implement this channel-truncated continuum theory concretely, we reuse the compactly supported smearings \eqref{eq:compactBumps} and evaluate \eqref{eq:eta1_oneparticle} numerically.  Fig.~\ref{fig:gappedOneParticleImplementation} compares three levels of description: the exact compact-support double integral within the one-particle truncation, the corresponding large distance center-to-center reduction \eqref{eq:eta1_c2c}--\eqref{eq:EB_oneparticle_c2c} obtained by evaluating the exact kernel at the support separation, and a faint pure-tail guide proportional to $r^{-2\gamma_j}e^{-2m_*r}$.  The solid and dashed curves therefore test the center-to-center reduction within the one-particle truncation. The figure provides a continuum implementation of the gapped current sector formula at the level of the lightest one-particle channel.

\subsection{Repulsive/BKT regime}
On the repulsive side relevant near the BKT line ($4\pi\le \beta_C^2<8\pi$), breathers are absent and the lightest neutral current channel is typically the soliton--antisoliton threshold, so that
\begin{equation}
M_{\rm eff}=2M
\qquad\text{(repulsive side, assuming the current first couples at the }s\bar s\text{ threshold)}.
\label{eq:Meff_repulsive}
\end{equation}
However, the threshold exponent $\gamma_j$ depends on the actual threshold behavior of the current form factor and is \emph{not} fixed here.  We therefore keep the repulsive/BKT asymptotic in the channel-dependent form
\begin{equation}
\eta_1(L,T)\sim r^{-\gamma_j}e^{-2Mr},
\qquad
E_B^{\max}(L,T)\sim r^{-2\gamma_j}e^{-4Mr},
\label{eq:repulsiveGammaUndetermined}
\end{equation}
with $\gamma_j$ left unspecified.

Near the BKT line, the SG mass scale has an essential singularity.  Because the RG flow depends on both the stiffness and the cosine amplitude, the distance to the critical separatrix is a nonuniversal combination of bare couplings, schematically
\begin{equation}
\delta\sim a_1(\beta_C^2-8\pi)+a_2\mu+\cdots,
\label{eq:delta_nonuniv}
\end{equation}
so that
\begin{equation}
M\sim \Lambda\exp\!\left(-\frac{c}{\sqrt{\delta}}\right),
\qquad
\xi_{\rm corr}^{(j)}\equiv\frac{1}{M_{\rm eff}}.
\label{eq:BKTcorr}
\end{equation}
Therefore the continuum QET envelope on the gapped side takes the form
\begin{equation}
E_B^{\max}(L,T)\sim r^{-2\gamma_j}\exp\!\left(-\frac{2r}{\xi_{\rm corr}^{(j)}}\right),
\qquad r=\sqrt{L^2-T^2},
\label{eq:BKTenvelope}
\end{equation}
in the large distance regime $s=r/\xi_{\rm corr}^{(j)}\gg 1$.  At fixed $r$, the approach to the BKT line eventually enters the crossover region $s\lesssim 1$, where the correct behavior is governed by the gapless law $E_B^{\max}\propto r^{-4}$ from Eq.~\eqref{eq:EBgaplessScaling}.  Equation~\eqref{eq:BKTenvelope} therefore describes the right tail on the gapped side, but not the full peak profile.

\begin{figure}[t]
  \centering
  \safeincludegraphics[width=0.78\linewidth]{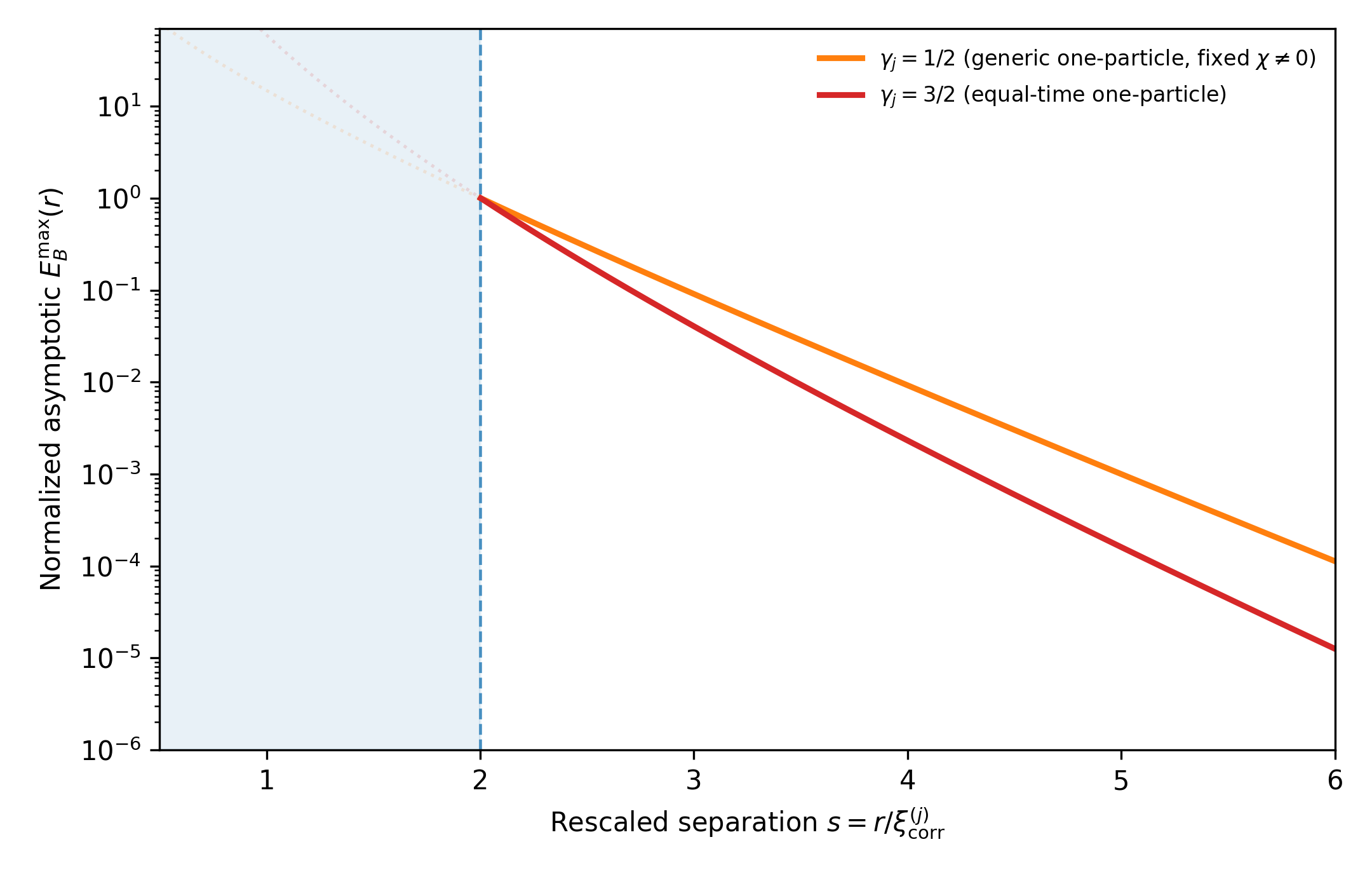}
  \caption{Normalized large distance behavior of the bulk gapped continuum signal for the one-particle channels whose exponents are fixed in the text.  The generic spacelike one-particle channel has $\gamma_j=1/2$, while the equal-time one-particle slice has $\gamma_j=3/2$.  The shaded region $s\lesssim 2$ indicates the crossover regime, and the asymptotic formulas apply for $s\gg 1$.  On the repulsive/BKT side the same exponential scale uses $M_{\rm eff}=2M$, but the threshold exponent depends on the channel and is therefore not plotted.}
  \label{fig:continuumAsymptotics}
\end{figure}

\section{Lattice analysis and connection to the open-chain lattice protocol}
\label{sec:latticeConnection}

\subsection{Structure of the lattice problem}
The lattice problem separates naturally into a neutral current sector and a charged sector.  The original single-site Pauli protocol of Ref.~\cite{Ikeda2023} probes only the charged part at separated sites, while a modified protocol on the same Hamiltonian isolates the neutral current sector and can be compared directly with the continuum current result.  The charged remainder of the original protocol is compared with a minimal boundary-state ansatz.  This keeps the exact neutral current statements separate from the more qualitative boundary discussion of the charged sector.

\subsection{The conventional single-site QET protocol}
\subsubsection{Weak-signal structure.}
A directly relevant lattice QET study is the open-chain massive-Thirring analysis of Ref.~\cite{Ikeda2023}.  There the qubit protocol is written in the generic Pauli form
\begin{equation}
P_{n_A}(\mu)=\frac12\bigl(\mathbf 1+\mu\sigma_{n_A}\bigr),
\qquad
U_{n_B}(\mu)=\cos\theta\,\mathbf 1-i\mu\sin\theta\,\sigma_{n_B},
\qquad
\mu=\pm 1,
\label{eq:IkedaGenericProtocol}
\end{equation}
with $\sigma_{n_A},\sigma_{n_B}\in\{X,Y,Z\}$.  In the concrete open-chain circuit shown in Fig.~1 of Ref.~\cite{Ikeda2023}, Alice uses $\sigma_{n_A}=X_{n_A}$ while Bob scans the three local Pauli axes at his site.

The exact extracted energy is
\begin{equation}
E_{B,{\rm lat}}=\frac12\left(\sqrt{\xi_{\rm lat}^2+\eta_{\rm lat}^2}-\xi_{\rm lat}\right),
\label{eq:EBlatExact}
\end{equation}
with
\begin{equation}
\xi_{\rm lat}=\expval{\sigma_{n_B}H\sigma_{n_B}}_g,
\qquad
\eta_{\rm lat}=\expval{\sigma_{n_A}\dot\sigma_{n_B}}_g,
\qquad
\dot\sigma_{n_B}=i[H,\sigma_{n_B}].
\label{eq:etalatdefs}
\end{equation}
Equivalently, before optimizing over $\theta$, Bob's local energy has the form
\begin{equation}
\expval{H_{n_B}}_{\rm lat}(\theta)
=\xi_{\rm lat}\sin^2\theta+\eta_{\rm lat}\cos\theta\sin\theta.
\label{eq:HBlatTheta}
\end{equation}
In the weak signal regime $|\eta_{\rm lat}|/\xi_{\rm lat}\ll1$ one obtains
\begin{equation}
\theta_*=-\frac{\eta_{\rm lat}}{2\xi_{\rm lat}}+O\!\left((\eta_{\rm lat}/\xi_{\rm lat})^3\right),
\qquad
E_{B,{\rm lat}}=\frac{\eta_{\rm lat}^2}{4\xi_{\rm lat}}+O\!\left(\frac{\eta_{\rm lat}^4}{\xi_{\rm lat}^3}\right).
\label{eq:EBlatWeak}
\end{equation}
Here Alice's measurement is projective.  The weak limit in Eq.~\eqref{eq:EBlatWeak} is a small-signal expansion of the optimized energy, not a weak measurement in the POVM sense.  This provides the structural point of contact with the continuum analysis.  The bosonized continuum protocol gives
\begin{equation}
\expval{H_B}_{\rm cont}(\kappa)=\frac12\xi\kappa^2+\eta\kappa,
\qquad
E_{B,{\rm cont}}^{\max}=\frac{\eta^2}{2\xi},
\label{eq:EBcontBridge}
\end{equation}
and hence at weak measurement
\begin{equation}
E_{B,{\rm cont}}^{\max}(\lambda_0)=\frac{\eta_1^2}{2\xi}\lambda_0^2+O(\lambda_0^4).
\label{eq:EBcontWeakBridge}
\end{equation}
Both descriptions therefore share the same quadratic structure, namely an Alice--Bob correlator squared divided by a local energetic cost.  The remaining question is which lattice operator sector corresponds to the continuum coefficient $\eta_1$.  For the original single-site protocol of Ref.~\cite{Ikeda2023}, the answer is not the neutral current sector, as shown below.  A direct continuum comparison therefore requires a separate neutral current construction on the same lattice Hamiltonian.

\subsubsection{Commutator decomposition.}
Write the open-chain spin Hamiltonian of Ref.~\cite{Ikeda2023} in the equivalent commutator-relevant form
\begin{equation}
H_{\rm lat}
=
-\frac{1}{4a}\sum_{n=1}^{N-1}\bigl(X_nX_{n+1}+Y_nY_{n+1}\bigr)
+\frac{m}{2}\sum_{n=1}^{N}\epsilon_n Z_n
+\frac{\Delta(g)}{4a}\sum_{n=1}^{N-1}\bigl(Z_nZ_{n+1}+Z_n+Z_{n+1}\bigr)
+\text{const},
\label{eq:IkedaHamiltonianComm}
\end{equation}
where $\epsilon_n=\pm1$ alternates along the chain and additive constants are omitted because they do not affect commutators.  For a bulk site $2\le n\le N-1$ the exact operator identities are
\begin{align}
\dot X_n
&=
-\frac{1}{2a}\bigl(Y_{n-1}Z_n+Z_nY_{n+1}\bigr)
-m\epsilon_n Y_n
-\frac{\Delta(g)}{2a}\bigl(Z_{n-1}Y_n+Y_nZ_{n+1}+2Y_n\bigr),
\label{eq:dotXIkeda}
\\
\dot Y_n
&=
+\frac{1}{2a}\bigl(X_{n-1}Z_n+Z_nX_{n+1}\bigr)
+m\epsilon_n X_n
+\frac{\Delta(g)}{2a}\bigl(Z_{n-1}X_n+X_nZ_{n+1}+2X_n\bigr),
\label{eq:dotYIkeda}
\\
\dot Z_n
&=
\frac{1}{2a}\bigl(Y_{n-1}X_n-X_{n-1}Y_n+X_nY_{n+1}-Y_nX_{n+1}\bigr).
\label{eq:dotZIkeda}
\end{align}
At the endpoints one must do slightly more than simply drop the nonexistent-neighbor terms: the $\Delta(g)$-induced on-site contribution is also halved because only one adjacent bond remains.  Explicitly,
\begin{align}
\dot X_1
&=
-\frac{1}{2a}Z_1Y_2
-m\epsilon_1 Y_1
-\frac{\Delta(g)}{2a}\bigl(Y_1Z_2+Y_1\bigr),
\nonumber\\
\dot Y_1
&=
+\frac{1}{2a}Z_1X_2
+m\epsilon_1 X_1
+\frac{\Delta(g)}{2a}\bigl(X_1Z_2+X_1\bigr),
\nonumber\\
\dot X_N
&=
-\frac{1}{2a}Y_{N-1}Z_N
-m\epsilon_N Y_N
-\frac{\Delta(g)}{2a}\bigl(Z_{N-1}Y_N+Y_N\bigr),
\nonumber\\
\dot Y_N
&=
+\frac{1}{2a}X_{N-1}Z_N
+m\epsilon_N X_N
+\frac{\Delta(g)}{2a}\bigl(Z_{N-1}X_N+X_N\bigr).
\label{eq:boundaryCommIkeda}
\end{align}
Equation~\eqref{eq:dotZIkeda} itself needs only the obvious deletion of the absent bond-current term at the boundary.
Defining the bond current with the same orientation as Table~II of Ref.~\cite{Ikeda2023},
\begin{equation}
\mathcal J_{n+1/2}^{\rm Ik}\equiv \frac{X_nY_{n+1}-Y_nX_{n+1}}{4a},
\label{eq:bondCurrentIkeda}
\end{equation}
Eq.~\eqref{eq:dotZIkeda} becomes
\begin{equation}
\dot Z_n=2\bigl(\mathcal J_{n+1/2}^{\rm Ik}-\mathcal J_{n-1/2}^{\rm Ik}\bigr).
\label{eq:dotZCurrentDifference}
\end{equation}

Specializing now to the concrete Alice choice $X_{n_A}$ of Ref.~\cite{Ikeda2023}, the three Bob channels decompose as
\begin{align}
\eta_{\rm lat}^{(X)}
&=
-\frac{1}{2a}\expval{X_{n_A}\bigl(Y_{n_B-1}Z_{n_B}+Z_{n_B}Y_{n_B+1}\bigr)}_g
-m\epsilon_{n_B}\expval{X_{n_A}Y_{n_B}}_g
\nonumber\\
&\qquad
-\frac{\Delta(g)}{2a}\expval{X_{n_A}\bigl(Z_{n_B-1}Y_{n_B}+Y_{n_B}Z_{n_B+1}+2Y_{n_B}\bigr)}_g
\nonumber\\
&\equiv
\eta_{\rm lat}^{(X;{\rm kin/str})}
+\eta_{\rm lat}^{(X;{\rm stag})}
+\eta_{\rm lat}^{(X;\rho{\rm -str})},
\label{eq:etaLatXdecomp}
\\[2mm]
\eta_{\rm lat}^{(Y)}
&=
+\frac{1}{2a}\expval{X_{n_A}\bigl(X_{n_B-1}Z_{n_B}+Z_{n_B}X_{n_B+1}\bigr)}_g
+m\epsilon_{n_B}\expval{X_{n_A}X_{n_B}}_g
\nonumber\\
&\qquad
+\frac{\Delta(g)}{2a}\expval{X_{n_A}\bigl(Z_{n_B-1}X_{n_B}+X_{n_B}Z_{n_B+1}+2X_{n_B}\bigr)}_g
\nonumber\\
&\equiv
\eta_{\rm lat}^{(Y;{\rm kin/str})}
+\eta_{\rm lat}^{(Y;{\rm stag})}
+\eta_{\rm lat}^{(Y;\rho{\rm -str})},
\label{eq:etaLatYdecomp}
\\[2mm]
\eta_{\rm lat}^{(Z)}
&=
2\expval{X_{n_A}\bigl(\mathcal J_{n_B+1/2}^{\rm Ik}-\mathcal J_{n_B-1/2}^{\rm Ik}\bigr)}_g
\equiv \eta_{\rm lat}^{(Z;j)}.
\label{eq:etaLatZdecomp}
\end{align}
The notation on the left indicates Bob's chosen control axis $\sigma_{n_B}=X_{n_B},Y_{n_B},Z_{n_B}$.  Equations~\eqref{eq:etaLatXdecomp}--\eqref{eq:etaLatZdecomp} are the interior-site formulas for $2\le n_B\le N-1$.  If Bob is placed at an endpoint, the boundary commutators \eqref{eq:boundaryCommIkeda} are used instead.

The sector content is now explicit.  Under the Jordan--Wigner map,
\begin{equation}
X_n=\mathcal S_{n-1}\bigl(\chi_n+\chi_n^\dagger\bigr),
\qquad
Y_n=-i\,\mathcal S_{n-1}\bigl(\chi_n-\chi_n^\dagger\bigr),
\qquad
\mathcal S_{n-1}\equiv \prod_{m<n}\bigl(1-2\chi_m^\dagger\chi_m\bigr),
\label{eq:JWstring}
\end{equation}
up to a convention-dependent phase.  Hence every term in \eqref{eq:etaLatXdecomp} and \eqref{eq:etaLatYdecomp} contains a nonlocal string/disorder factor on Bob's side, as well as either the explicit staggered mass factor $m\epsilon_{n_B}$ or density dressing through the neighboring $Z$ operators.  By contrast, the $Z$ channel \eqref{eq:etaLatZdecomp} is a pure local current difference on Bob's side.  At the operator level, the conserved current sector appears cleanly only in the Bob-side $Z$ channel, whereas the $X/Y$ channels remain charged string/disorder sectors.

\subsubsection{\texorpdfstring{$U(1)$ selection rule: why the actual current channel vanishes.}{U(1) selection rule: why the actual current channel vanishes.}}
The previous subsection isolates the neutral current channel on Bob's side, but for the concrete Alice choice $X_{n_A}$ of Ref.~\cite{Ikeda2023} that neutral channel does \emph{not} survive inside the actual weak signal.  The reason is the exact lattice $U(1)$ symmetry generated by the total fermion number
\begin{equation}
Q\equiv \sum_{n=1}^{N} n_n = \frac12\sum_{n=1}^{N}(Z_n+1),
\qquad [H_{\rm lat},Q]=0.
\label{eq:U1charge}
\end{equation}
Introduce raising/lowering operators
\begin{equation}
\sigma_n^\pm\equiv \frac{X_n\pm iY_n}{2},
\qquad
X_n=\sigma_n^++\sigma_n^-,
\qquad
Y_n=-i\bigl(\sigma_n^+-\sigma_n^-\bigr).
\label{eq:spinLadder}
\end{equation}
Then
\begin{equation}
\mathcal J_{n+1/2}^{\rm Ik}
=\frac{i}{2a}\bigl(\sigma_n^+\sigma_{n+1}^- - \sigma_n^-\sigma_{n+1}^+\bigr),
\qquad
[Q,\mathcal J_{n+1/2}^{\rm Ik}]=[Q,Z_n]=0,
\label{eq:currentNeutral}
\end{equation}
so both the bond current and $Z_n$ are neutral under the conserved charge.  By contrast, $X_n$ and $Y_n$ form a charged doublet:
\begin{equation}
e^{i\alpha Q}
\begin{pmatrix}
X_n\\ Y_n
\end{pmatrix}
e^{-i\alpha Q}
=
\begin{pmatrix}
\cos\alpha & -\sin\alpha\\
\sin\alpha & \cos\alpha
\end{pmatrix}
\begin{pmatrix}
X_n\\ Y_n
\end{pmatrix}.
\label{eq:XYrotation}
\end{equation}
On a finite open chain the ground state used in exact diagonalization can be chosen as a simultaneous eigenstate $|g_q\rangle$ of $H_{\rm lat}$ and $Q$.  Therefore, for any neutral operator $O_{\rm neut}$ with $[Q,O_{\rm neut}]=0$, the vector
\begin{equation}
\bm v_n(O_{\rm neut})
\equiv
\begin{pmatrix}
\expval{X_n O_{\rm neut}}_{g_q}\\[2pt]
\expval{Y_n O_{\rm neut}}_{g_q}
\end{pmatrix}
\end{equation}
must satisfy $\bm v_n(O_{\rm neut})=R(\alpha)\bm v_n(O_{\rm neut})$ for every rotation matrix $R(\alpha)$ from \eqref{eq:XYrotation}.  The only such vector is the zero vector, so
\begin{equation}
\expval{X_n O_{\rm neut}}_{g_q}=\expval{Y_n O_{\rm neut}}_{g_q}=0
\qquad
\text{for every neutral }O_{\rm neut}.
\label{eq:neutralSelectionRule}
\end{equation}
Applying this to $O_{\rm neut}=\mathcal J_{n_B+1/2}^{\rm Ik}-\mathcal J_{n_B-1/2}^{\rm Ik}$ gives the exact finite chain result
\begin{equation}
\eta_{\rm lat}^{(Z)}=0
\qquad
\text{for the concrete Alice choice }X_{n_A}.
\label{eq:etaLatZvanishes}
\end{equation}
The neutral smooth-current sector therefore does \emph{not} contribute to the actual weak signal when Alice uses $X_{n_A}$.  Hence any nonzero actual weak signal must reside in the charged $X/Y$ channels, namely in the string/disorder and staggered/density sectors exhibited explicitly in \eqref{eq:etaLatXdecomp} and \eqref{eq:etaLatYdecomp}.  Determining which of those charged sectors dominates is a separate dynamical question.

\begin{figure}[t]
  \centering
  \safeincludegraphics[width=0.72\linewidth]{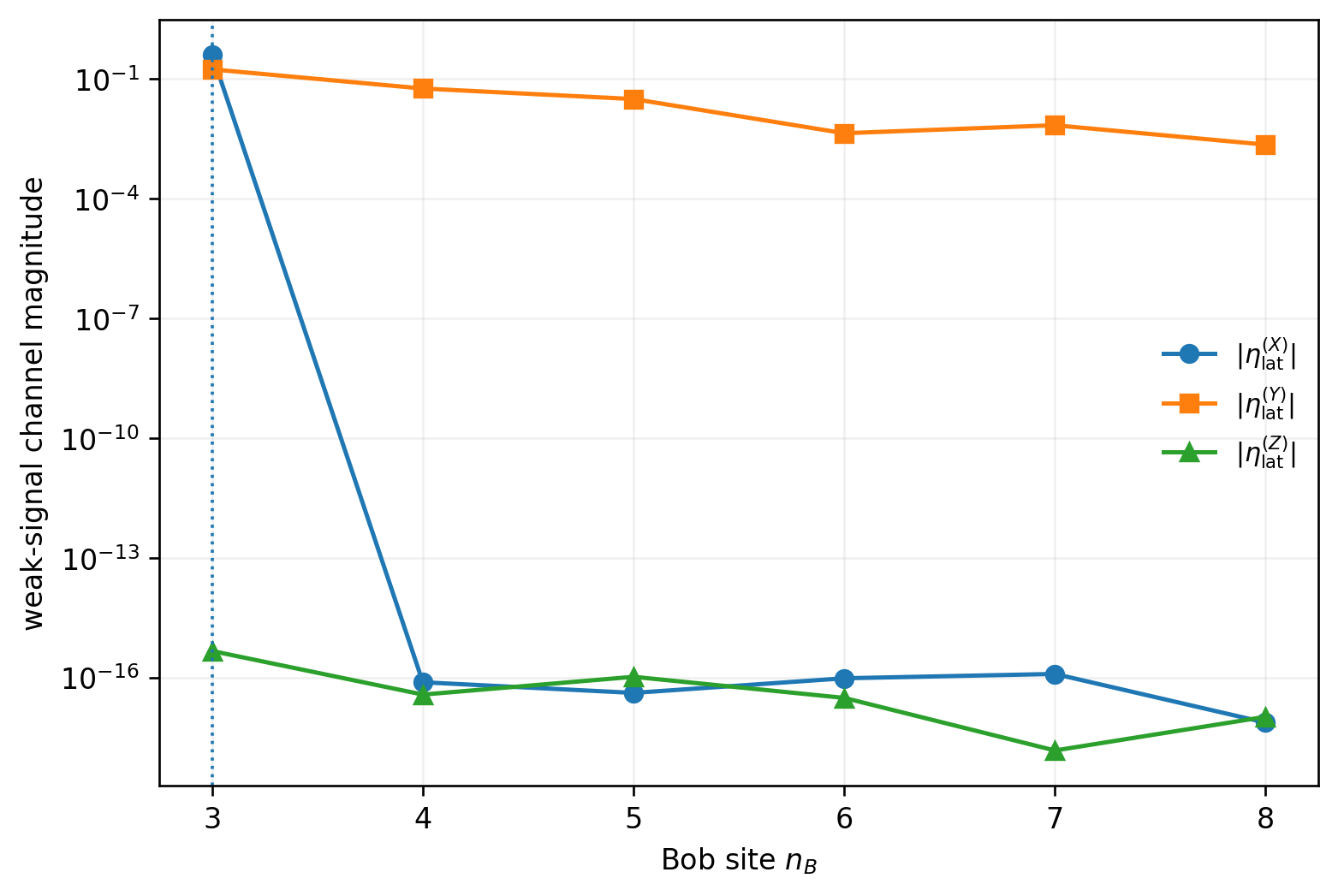}
  \caption{Exact-diagonalization illustration of the $U(1)$ selection rule on an interacting open chain.  For the concrete Alice choice $X_{n_A}$ used in Ref.~\cite{Ikeda2023}, the Bob-side $Z$ channel of the actual weak signal is symmetry-forbidden and remains at machine precision.  In the displayed example the separated nonzero signal is carried predominantly by the charged $Y$ channel, while the charged $X$ channel survives only at the short distance contact point $n_B=n_A+1$.  The example shown uses $N=8$, $m=0.4$, $\Delta(g)=0.3$, and Alice at $n_A=2$.}
  \label{fig:ikedaSelectionRule}
\end{figure}

Fig.~\ref{fig:ikedaSelectionRule} illustrates \eqref{eq:etaLatZvanishes} numerically on a small interacting open chain.  The Bob-side $Z$ channel sits at machine precision throughout, whereas the surviving separated signal lies predominantly in the charged $Y$ channel.  In the displayed example the $X$ channel survives only at the contact point $n_B=n_A+1$ and is otherwise strongly suppressed.  The next subsection sharpens that observation into an exact real basis suppression statement for separated Bob sites.

\subsubsection{\texorpdfstring{Real-basis constraint: the separated signal is effectively a $Y$ channel.}{Real-basis constraint: the separated signal is effectively a Y channel.}}
The exact $U(1)$ null result still leaves the charged $X/Y$ channels.  On a finite open chain in the computational $\sigma^z$ basis, the lattice Hamiltonian is real symmetric, and for the nondegenerate finite-chain ground states used below one may choose the ground-state vector real. Only under these finite-chain, real-basis, and nondegenerate-ground-state assumptions can one sharpen the statement further. For Bob sites separated from Alice,
\begin{equation}
n_B\ge n_A+2,
\label{eq:separatedSites}
\end{equation}
Eq.~\eqref{eq:dotXIkeda} shows that $\dot X_{n_B}$ contains exactly one $Y$ matrix and otherwise only real matrices, hence it is purely imaginary in that basis.  Since its support is now disjoint from $X_{n_A}$, the product $X_{n_A}\dot X_{n_B}$ is both Hermitian and purely imaginary.  Therefore,
\begin{equation}
\eta_{\rm lat}^{(X)}(n_A,n_B)
=\expval{X_{n_A}\,\dot X_{n_B}}_g
=0,
\qquad n_B\ge n_A+2,
\label{eq:etaLatXSeparatedVanish}
\end{equation}
in the real ground-state convention.  By contrast, Eq.~\eqref{eq:dotYIkeda} is real, so $X_{n_A}\dot Y_{n_B}$ is real Hermitian and can remain nonzero at separated sites.  Thus, in the displayed open-chain geometries the actual separated signal is effectively a Bob-side $Y$-channel signal, while the Bob-side $X$ channel survives only as a contact or very short distance effect.  Accordingly, the direct boundary comparison of the charged sector is carried out for $\eta_Y$ rather than for $\eta_X$.

The original protocol therefore has a clear sector structure.  The neutral current sector is absent, and the surviving separated signal lies in a charged $Y$ channel.  A direct comparison with the continuum current result therefore requires a distinct neutral current protocol on the same lattice Hamiltonian.

\subsection{Neutral-current lattice realization}
\subsubsection{Current sector on the lattice.}
Because Eq.~\eqref{eq:etaLatZvanishes} removes the Bob-side neutral current from the original Alice-$X$ protocol, any direct comparison with the continuum current sector must use a separate neutral observable defined on the same lattice Hamiltonian.  It can be formulated either as a coarse grained current correlator or as a modified lattice protocol whose weak signal is exactly that correlator.

Define the lattice density
\begin{equation}
n_n\equiv \frac{Z_n+1}{2}.
\label{eq:latticeDensity}
\end{equation}
Equation~\eqref{eq:dotZCurrentDifference} implies the exact discrete continuity equation
\begin{equation}
\dot n_n+\mathcal J_{n-1/2}^{\rm Ik}-\mathcal J_{n+1/2}^{\rm Ik}=0,
\label{eq:latticeContinuity}
\end{equation}
which fixes \eqref{eq:bondCurrentIkeda} as the natural local current up to the trivial choice of spatial orientation.  This is the correct starting point for a continuum current comparison.

\subsubsection{Modified neutral current protocol.}
\label{sec:modifiedNeutralBridge}
A neutral current construction on the same lattice Hamiltonian is defined through coarse grained lattice currents centered at Alice and Bob,
\begin{equation}
J_{{\rm cg},\ell}^{(A)}=\sum_{m=1}^{N-1}f_{A,\ell}(x_{m+1/2})\,\mathcal J_{m+1/2}^{\rm Ik},
\qquad
J_{{\rm cg},\ell}^{(B)}=\sum_{n=1}^{N-1}g_{B,\ell}(x_{n+1/2})\,\mathcal J_{n+1/2}^{\rm Ik},
\label{eq:coarseCurrents}
\end{equation}
with smooth envelopes of width $\ell$ chosen in the scaling window
\begin{equation}
a\ll \ell \ll \xi_{\rm corr}^{(j)}.
\label{eq:scalingWindow}
\end{equation}
The associated same-current sector lattice observable is
\begin{equation}
\eta_{JJ,\ell}^{\rm lat}(L,T)\equiv
2\expval{J_{{\rm cg},\ell}^{(A)}(0)\,J_{{\rm cg},\ell}^{(B)}(T)}_g.
\label{eq:etaJJproxy}
\end{equation}
This is \emph{not} a sector decomposition of the actual Alice-$X$ weak signal.  It is a separately defined neutral sector observable on the same massive-Thirring lattice Hamiltonian.

The same quantity is also the weak signal functional of a modified neutral protocol on the same lattice Hamiltonian.  Let Alice use the lattice analog of the two-outcome current POVM of Refs.~\cite{Hotta2008,HottaEM2010},
\begin{equation}
M_A^{(J)}(\mu)=\frac{1}{\sqrt2}\Bigl[\cos\bigl(\lambda_0 J_{{\rm cg},\ell}^{(A)}\bigr)+\mu\sin\bigl(\lambda_0 J_{{\rm cg},\ell}^{(A)}\bigr)\Bigr],
\qquad \mu=\pm1,
\label{eq:latticeCurrentPOVM}
\end{equation}
and let Bob use a bounded neutral unitary generated by a localized smeared $Z$ operator,
\begin{equation}
U_B^{(h)}(\mu)=\exp\!\bigl[-i\mu\theta\,O_{B,h}\bigr],
\qquad
O_{B,h}\equiv \sum_{n=1}^{N} h_{B,\ell}(x_n) Z_n.
\label{eq:latticeNeutralBobUnitary}
\end{equation}
At leading order in the measurement strength the corresponding lattice weak signal is
\begin{equation}
\eta_{JZ,\ell}^{\rm lat}[h_B](T)
\equiv
\expval{J_{{\rm cg},\ell}^{(A)}(0)\,\dot O_{B,h}(T)}_g.
\label{eq:etaJZdef}
\end{equation}
Using \eqref{eq:dotZCurrentDifference} and the open-chain relations $\mathcal J_{1/2}^{\rm Ik}=\mathcal J_{N+1/2}^{\rm Ik}=0$, one finds the exact discrete integration-by-parts identity
\begin{equation}
\dot O_{B,h}(T)
=2\sum_{n=1}^{N-1}\Bigl[h_{B,\ell}(x_n)-h_{B,\ell}(x_{n+1})\Bigr]\,\mathcal J_{n+1/2}^{\rm Ik}(T).
\label{eq:discreteIBP}
\end{equation}
Therefore, if Bob's bond-current smearing is chosen as
\begin{equation}
g_{B,\ell}(x_{n+1/2})\equiv h_{B,\ell}(x_n)-h_{B,\ell}(x_{n+1}),
\label{eq:gFromh}
\end{equation}
then the modified neutral weak signal reduces exactly to
\begin{equation}
\eta_{JZ,\ell}^{\rm lat}[h_B](L,T)=2\expval{J_{{\rm cg},\ell}^{(A)}(0)\,J_{{\rm cg},\ell}^{(B)}(T)}_g
=\eta_{JJ,\ell}^{\rm lat}(L,T).
\label{eq:etaJZequalsJJ}
\end{equation}
The same lattice Hamiltonian thus supports a neutral current protocol whose weak signal is exactly the corresponding current correlator.  This is the lattice sector compared with the continuum current result.

\begin{figure}[t]
  \centering
  \safeincludegraphics[width=0.98\linewidth]{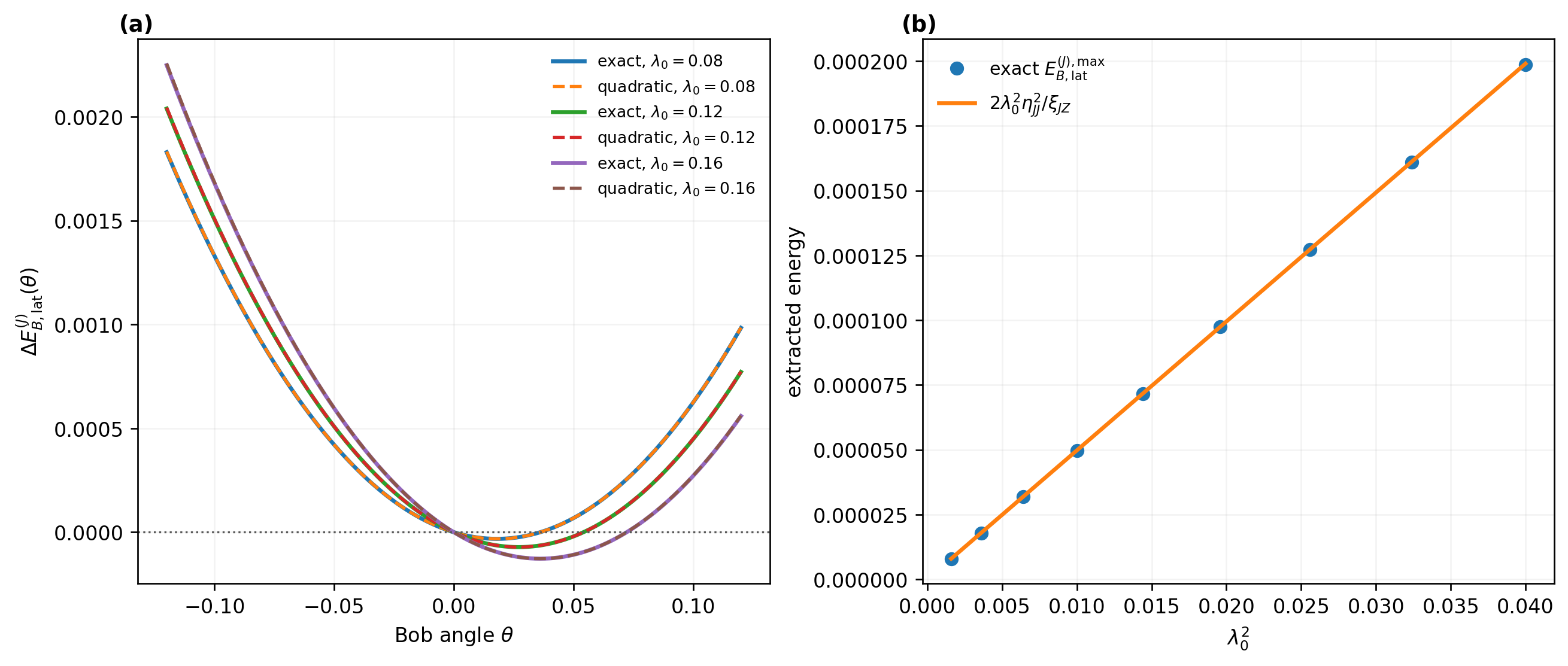}
  \caption{Neutral-current protocol on the same interacting open chain ($N=8$, $m=0.4$, $\Delta(g)=0.3$).  (a) Bob-induced energy change $\Delta E_{B,{\rm lat}}^{(J)}(\theta)$ for three weak measurement strengths $\lambda_0$, compared with the quadratic prediction \eqref{eq:modifiedEnergyQuadratic}.  (b) extracted energy $E_{B,{\rm lat}}^{(J),\max}=-\min_\theta \Delta E_{B,{\rm lat}}^{(J)}(\theta)$ as a function of $\lambda_0^2$, with slope $2[\eta_{JJ,\ell}^{\rm lat}]^2/\xi_{JZ,\ell}^{\rm lat}$.  The agreement verifies the leading weak measurement QET law for the lattice current sector.}
  \label{fig:modifiedNeutralProtocol}
\end{figure}

To complete the modified neutral construction as a lattice QET protocol, one also needs its energetic cost.  Write the kinetic bond operator
\begin{equation}
K_{n+1/2}\equiv -\frac{X_nX_{n+1}+Y_nY_{n+1}}{4a},
\label{eq:kineticBond}
\end{equation}
so that $H_{\rm lat}=\sum_{n=1}^{N-1}K_{n+1/2}+H_Z+\text{const}$, where $H_Z$ contains the staggered-mass and density terms and commutes with $O_{B,h}$.  Because $O_{B,h}$ generates site-dependent $Z$ rotations,
\begin{equation}
U_B^{(h)}(\mu)^\dagger \sigma_n^\pm U_B^{(h)}(\mu)=e^{\pm 2 i \mu \theta h_{B,\ell}(x_n)}\sigma_n^\pm,
\label{eq:localZrotation}
\end{equation}
one obtains the exact bondwise conjugation formula
\begin{equation}
U_B^{(h)}(\mu)^\dagger K_{n+1/2}U_B^{(h)}(\mu)
=
K_{n+1/2}\cos\!\bigl(2\mu\theta g_{B,\ell}(x_{n+1/2})\bigr)
-
\mathcal J_{n+1/2}^{\rm Ik}\sin\!\bigl(2\mu\theta g_{B,\ell}(x_{n+1/2})\bigr).
\label{eq:bondRotation}
\end{equation}
Thus only the bonds with nonzero $g_{B,\ell}$ contribute to Bob's local energy change.  At Bob time $T$, the post-feedback energy change relative to the post-Alice state obeys the weak measurement expansion
\begin{equation}
\Delta E_{B,{\rm lat}}^{(J)}(\theta;L,T)
=
-\theta\,\expval{D_A^{(J)}\,\dot O_{B,h}(T)}_g
+\frac{\theta^2}{2}\,\xi_{JZ,\ell}^{\rm lat}[h_B]
+O(\lambda_0^3\theta,\lambda_0^2\theta^2,\theta^3),
\label{eq:modifiedEnergyExpansion}
\end{equation}
where
\begin{equation}
D_A^{(J)}\equiv \sum_{\mu=\pm1}\mu\,M_A^{(J)}(\mu)^\dagger M_A^{(J)}(\mu)
=\sin\!\bigl(2\lambda_0 J_{{\rm cg},\ell}^{(A)}\bigr),
\label{eq:DAJ}
\end{equation}
and
\begin{equation}
\xi_{JZ,\ell}^{\rm lat}[h_B]
\equiv
\expval{[O_{B,h},[H_{\rm lat},O_{B,h}]]}_g
=
-4\sum_{n=1}^{N-1} g_{B,\ell}(x_{n+1/2})^2 \expval{K_{n+1/2}}_g \ge 0.
\label{eq:xiJZdef}
\end{equation}
Using $D_A^{(J)}=2\lambda_0 J_{{\rm cg},\ell}^{(A)}+O(\lambda_0^3)$ as well as \eqref{eq:etaJZequalsJJ}, the leading weak measurement lattice QET law becomes
\begin{equation}
\Delta E_{B,{\rm lat}}^{(J)}(\theta;L,T)
=
-2\lambda_0\theta\,\eta_{JJ,\ell}^{\rm lat}(L,T)
+\frac{\theta^2}{2}\,\xi_{JZ,\ell}^{\rm lat}[h_B]
+O(\lambda_0^3\theta,\lambda_0^2\theta^2,\theta^3).
\label{eq:modifiedEnergyQuadratic}
\end{equation}
Minimizing over $\theta$ yields
\begin{equation}
\theta_*^{(J)}=
\frac{2\lambda_0\,\eta_{JJ,\ell}^{\rm lat}(L,T)}{\xi_{JZ,\ell}^{\rm lat}[h_B]}
+O(\lambda_0^3),
\qquad
E_{B,{\rm lat}}^{(J),\max}
=
\frac{2\lambda_0^2\,[\eta_{JJ,\ell}^{\rm lat}(L,T)]^2}{\xi_{JZ,\ell}^{\rm lat}[h_B]}
+O(\lambda_0^4).
\label{eq:modifiedEnergyOptimized}
\end{equation}
The overall factor $2$ reflects the convention $O_{B,h}=\sum h Z$.  Rescaling the generator by $1/2$ would absorb it.  The neutral current construction is therefore a lattice QET protocol at leading order, with the same quadratic structure as the continuum analysis.  Fig.~\ref{fig:modifiedNeutralProtocol} checks this directly on a small interacting open chain.

In the continuum scaling window one expects
\begin{equation}
\eta_{JJ,\ell}^{\rm lat}(L,T)
=
C_{\rm lat}^{(J)}(\ell;x_*,y_*)\,
\frac{\beta_C^2}{8\pi^2}\,
\eta_1(L,T)
+\text{scaling corrections},
\label{eq:etaJJbridge}
\end{equation}
but the normalization factor is \emph{not} left free.  It is fixed by a one-point equal-time current correlator matching condition,
\begin{equation}
C_{\rm lat}^{(J)}(\ell;x_*,y_*)
\equiv
\frac{\Gamma_{\ell}^{\rm lat}(x_*,y_*)}{\Gamma_{\ell}^{\rm cont,cand}(x_*,y_*)},
\qquad
\Gamma_{\ell}^{\rm lat}(x,y)\equiv
\expval{J_{{\rm cg},\ell}(0,x)\,J_{{\rm cg},\ell}(0,y)}_g,
\label{eq:ClatFix}
\end{equation}
with
\begin{equation}
\Gamma_{\ell}^{\rm cont,cand}(x,y)
\equiv
\int du\,dv\,
\phi_\ell(u-x)\phi_\ell(v-y)
\expval{j^1(0,u)j^1(0,v)}_{\rm cont,cand}.
\label{eq:GammaContCand}
\end{equation}
Equation~\eqref{eq:etaJJbridge} gives the continuum--lattice relation for the neutral current sector.  For the numerical test we fix the coupling, the open-chain geometry, the smoothing prescription, and the time slice, and work at the free open-chain point $m=\Delta(g)=0$.  After one-point normalization, the remaining distance dependence is parameter-free within that setting.

\subsubsection{Free-point boundary test.}
Because Ref.~\cite{Ikeda2023} uses an open chain, the relevant continuum boundary condition must be tied to the lattice geometry.  The exact lattice statement is simply that there is no bond outside the chain, so the external current vanishes.  The continuum counterpart is therefore the no-flux condition
\begin{equation}
j^1(t,x)\big|_{x\in \partial I}=0.
\label{eq:noFluxBoundary}
\end{equation}
Using the bosonization map
\begin{equation}
j^1=\frac{\beta_C}{2\pi}\Pi=\frac{\beta_C}{2\pi}\partial_t\phi,
\label{eq:currentMomentumBoundary}
\end{equation}
the most natural bosonic realization is Dirichlet for $\phi$ (equivalently Neumann for the dual field), for which the equal-time finite-interval reference kernel is
\begin{equation}
\expval{j^1(0,x)j^1(0,y)}_{{\rm cand},D}
\propto
\sum_{m\in\mathbb Z}
\left[
\frac{1}{(x-y+2mL_{\rm sys})^2}
-
\frac{1}{(x+y+2mL_{\rm sys})^2}
\right].
\label{eq:DirichletCurrentKernel}
\end{equation}
The Neumann kernel has the opposite sign in front of the image term.  Because the interacting gapped boundary SG correlator is not computed here, \eqref{eq:DirichletCurrentKernel} is used as a boundary-dependent continuum reference kernel for the neutral sector.

\begin{figure}[t]
  \centering
  \safeincludegraphics[width=0.98\linewidth]{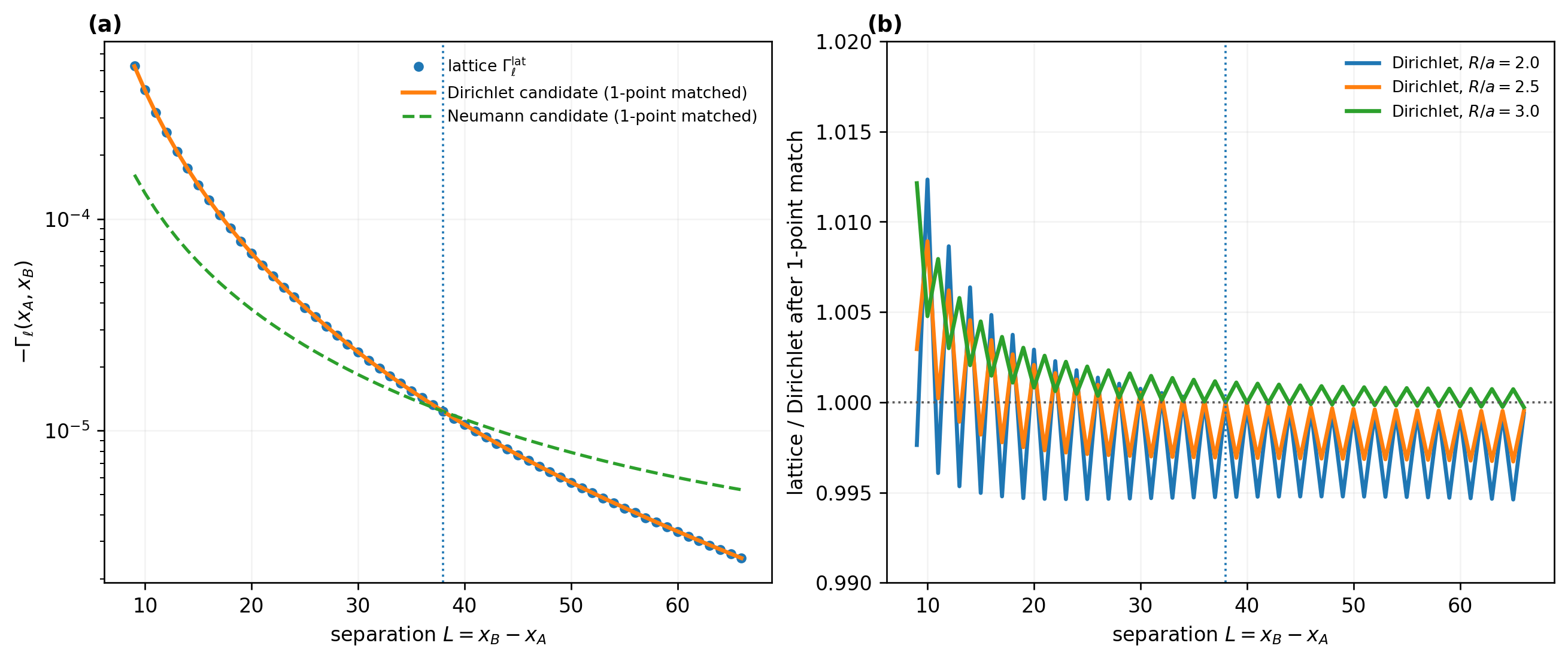}
  \caption{Free-point boundary test for the neutral current sector at $m=\Delta(g)=0$.  (a) Lattice data for the coarse grained bond-current correlator $\Gamma_{\ell}^{\rm lat}(x_A,x_B)$ on an open chain, compared with Dirichlet and Neumann finite-interval image kernels normalized at a single reference point through \eqref{eq:ClatFix}.  (b) After the same one-point matching, the Dirichlet ratio stays close to unity for $R/a=2.0,2.5,3.0$.  The same conclusion holds when the reference point is changed, which supports a no-flux/Dirichlet interpretation for the neutral conserved current sector at the gapless free point.}
  \label{fig:latticeCurrentBridge}
\end{figure}

We test this same current sector numerically at the gapless free open-chain point $m=\Delta(g)=0$.  Panel~(a) of Fig.~\ref{fig:latticeCurrentBridge} compares the one-point matched lattice current correlator with the Dirichlet and Neumann image-kernel references.  The Dirichlet kernel follows the lattice profile closely, whereas the Neumann kernel does not.  Panel~(b) shows that this agreement survives changes of the coarse-graining width: for $R/a\in\{2.0,2.5,3.0\}$ the one-point matched Dirichlet ratios remain within a few percent of unity, while the corresponding Neumann ratios remain order-one away.  Moving the normalization point from $L_*=19a$ to $38a$ to $49a$ does not change this conclusion.  The free-point comparison therefore favors the Dirichlet image kernel as the continuum reference for the neutral conserved current sector.

With the neutral current relation in place, the actual nonzero weak signal of the Alice-$X$ circuit lies in the charged sector.  That sector is analyzed at the operator level and then compared with a boundary-state sign test.

\subsection{Charged sector and boundary comparison}
\subsubsection{Charged scaling fields.}
The selection rule shows that the next step toward the actual minimal qubit protocol of Ref.~\cite{Ikeda2023} is the bosonization of the surviving charged $X/Y$ sector itself.  The natural objects are the ladder operators $\sigma_l^\pm$.  In the critical XXZ/Thirring scaling limit, Ref.~\cite{LukyanovTerras2003} gives the expansion in semi-local Gaussian fields as
\begin{align}
\sigma_l^\pm
&\sim
C_0 a^{d_{1,0}}\,\mathcal O_{\pm1,0}(x)
+\frac{(-1)^l C_1}{2}\,a^{d_{1,2}}\Bigl(\mathcal O_{\pm1,2}(x)+\mathcal O_{\pm1,-2}(x)\Bigr)
+\cdots,
\label{eq:LTsigmaPM}
\\
\sigma_l^z
&\sim
\frac{a}{2\pi\sqrt{\eta_{\rm XXZ}}}\,\partial_t\varphi(x)
+\frac{(-1)^l C_1^z}{2i}\,a^{d_{0,2}}\Bigl(\mathcal O_{0,2}(x)-\mathcal O_{0,-2}(x)\Bigr)
+\cdots,
\label{eq:LTsigmaZ}
\end{align}
with $x=la$ and
\begin{equation}
\begin{aligned}
\mathcal O_{s,n}(x,t)
={}&\Lambda^{d_{s,n}}
\lim_{\epsilon\to+0}
\exp\!\left[\frac{i n}{4\sqrt{\eta_{\rm XXZ}}}\int_{-\infty}^{x}dx'\,\partial_t\varphi(x',t)\right]
\\
&\times
\exp\!\left[\frac{i s\sqrt{\eta_{\rm XXZ}}}{2}\,\varphi(x+\epsilon,t)\right],
\qquad
 d_{s,n}=\frac{s^2\eta_{\rm XXZ}}{2}+\frac{n^2}{8\eta_{\rm XXZ}}.
\end{aligned}
\label{eq:LTfields}
\end{equation}
The first exponential is precisely the disorder / Jordan--Wigner string, whereas the second is a local vertex operator.  Under the lattice $U(1)$ symmetry these fields carry charge $s$,
\begin{equation}
U_\alpha\,\mathcal O_{s,n}\,U_\alpha^{-1}=e^{is\alpha}\mathcal O_{s,n},
\label{eq:LTU1}
\end{equation}
so $\mathcal O_{\pm1,0}$ are the leading charged continuum avatars of $\sigma^\pm$.  In the SG language they are the topologically charged dual exponentials / Mandelstam operators that create one-soliton charge~\cite{Mandelstam1975,LukyanovZamolodchikov2001}, while $\mathcal O_{\pm1,\pm2}$ are the first charged harmonics obtained by dressing them with the neutral $\sigma^z$ harmonic.

To avoid any clash with the QET weak signal coefficient $\eta(\lambda_0)$, we denote the Lukyanov--Terras XXZ parameter by $\eta_{\rm XXZ}$.  In the critical XXZ window $-1<\Delta_{\rm XXZ}<1$ one may write
\begin{equation}
\Delta_{\rm XXZ}=-\cos(\pi\eta_{\rm XXZ}),
\qquad
K=\frac{1}{2\eta_{\rm XXZ}},
\qquad
\beta_C^2=8\pi\eta_{\rm XXZ},
\label{eq:etaKbetaMap}
\end{equation}
where $K$ is the Luttinger parameter and $\beta_C$ is the canonical SG coupling used in the continuum sections.  In this notation, $\eta_{\rm XXZ}$ fixes the charged UV power, $K=(2\eta_{\rm XXZ})^{-1}$, $\beta_C^2=8\pi\eta_{\rm XXZ}$, and $\Delta(g)$ is used below only as the lattice-side interaction parameter in the finite-mass examples.  Table~\ref{tab:chargedParameterMap} summarizes the notation so that the charged formulas can be read without repeatedly translating between the XXZ and SG conventions.
\begin{table}[t]
\centering
\footnotesize
\setlength{\tabcolsep}{4pt}
\begin{tabular}{|P{0.18\linewidth}|P{0.30\linewidth}|P{0.37\linewidth}|}
\hline
Symbol & Meaning in this paper & Relation used here \\
\hline
$\eta_{\rm XXZ}$ & Lukyanov--Terras scaling-field parameter & $\Delta_{\rm XXZ}=-\cos(\pi\eta_{\rm XXZ})$, $K=(2\eta_{\rm XXZ})^{-1}$ \\
\hline
$K$ & Luttinger parameter of the critical XXZ/Thirring scaling limit & $\beta_C^2=4\pi/K=8\pi\eta_{\rm XXZ}$ \\
\hline
$\beta_C$ & Canonical SG coupling used in Secs.~2--5 & same coupling that appears in the bosonized current map \\
\hline
$\Delta_{\rm XXZ}$ & Massless XXZ anisotropy entering the scaling-field dictionary & identified with the lattice coefficient $\Delta(g)$ only in the massless, staggered-rotated dictionary \\
\hline
$\Delta(g)$ & Interaction coefficient of the open-chain lattice Hamiltonian used numerically & treated only as the lattice-side parameter whose long distance charged sector is organized by the XXZ dictionary \\
\hline
\end{tabular}
\caption{Parameter dictionary used in the charged sector.  For the finite-mass, finite-size open-chain examples below, this is a dictionary of leading long distance sectors, not an exact quantitative finite-size map.}
\label{tab:chargedParameterMap}
\end{table}
For the massless limit $m=0$ of the open-chain lattice Hamiltonian, a staggered $\pi$ rotation about $Z$ flips the sign of the XY term and identifies the coefficient $\Delta(g)$ with the usual XXZ anisotropy $\Delta_{\rm XXZ}$ in this scaling-field dictionary.  For the finite-mass, finite-size open-chain examples below, Eq.~\eqref{eq:etaKbetaMap} and Table~\ref{tab:chargedParameterMap} are used only to organize the leading scaling fields and not as an exact quantitative map.  In particular,
\begin{equation}
d_{1,0}=\frac{\eta_{\rm XXZ}}{2},
\qquad
\expval{\mathcal O_{+1,0}(r)\mathcal O_{-1,0}(0)}\propto r^{-\eta_{\rm XXZ}}
\quad (Mr\ll1),
\label{eq:chargedUVScaling}
\end{equation}
so the charged critical power is $\eta_{\rm XXZ}$-dependent even though the massive infrared tail used later is universal.

\subsubsection{Ultraviolet and infrared regimes.} The UV law $r^{-\eta_{\rm XXZ}}$ and the infrared one-soliton tail $r^{-1/2}e^{-Mr}$ belong to different regimes.  The comparison here uses only the infrared regime.  Interpolating between the two, including the exact boundary normalization, is not attempted.

Equations~\eqref{eq:LTsigmaPM}--\eqref{eq:LTfields} give a bosonized description of the actual nonzero weak signal.  Define the opposite-charge pairings
\begin{equation}
C_{+-}(n_A,n_B)\equiv \expval{\sigma_{n_A}^+\,\dot\sigma_{n_B}^-}_g,
\qquad
C_{-+}(n_A,n_B)\equiv \expval{\sigma_{n_A}^-\,\dot\sigma_{n_B}^+}_g.
\label{eq:chargedPairings}
\end{equation}
Because the ground state has definite total charge, the same-sign pairings vanish exactly,
\begin{equation}
\expval{\sigma_{n_A}^+\,\dot\sigma_{n_B}^+}_g
=
\expval{\sigma_{n_A}^-\,\dot\sigma_{n_B}^-}_g
=0,
\label{eq:sameSignChargedVanish}
\end{equation}
so the actual Bob-side $X/Y$ channels reduce to the exact identities
\begin{equation}
\eta_{\rm lat}^{(X)}=C_{+-}+C_{-+},
\qquad
\eta_{\rm lat}^{(Y)}=i\bigl(C_{+-}-C_{-+}\bigr).
\label{eq:etaXYfromChargedPairs}
\end{equation}
Combined with the separated-site suppression \eqref{eq:etaLatXSeparatedVanish}, this means that for $n_B\ge n_A+2$ the actual weak signal is encoded in a \emph{single} opposite-charge pairing,
\begin{equation}
C_{-+}(n_A,n_B)=-C_{+-}(n_A,n_B),
\qquad
\eta_{\rm lat}^{(Y)}(n_A,n_B)=2i\,C_{+-}(n_A,n_B).
\label{eq:etaYSinglePair}
\end{equation}
Thus the entire nonzero separated part of the actual Alice-$X$ weak signal is encoded in oppositely charged pairings of semi-local fields.  The neutral current result of Sec.~\ref{sec:latticeConnection} therefore identifies the charged correlators that replace the current correlator in the continuum description of the original protocol.

At leading continuum order the correspondence with the exact lattice decomposition \eqref{eq:etaLatXdecomp}--\eqref{eq:etaLatYdecomp} is direct.  The kinetic/string piece $\eta_{\rm lat}^{(Y;{\rm kin/str})}$ is the descendant of the leading charged field $\mathcal O_{\pm1,0}$, the explicit staggered term $\eta_{\rm lat}^{(Y;{\rm stag})}$ multiplies that same charged field by the continuum mass perturbation, and the density-dressed string piece $\eta_{\rm lat}^{(Y;\rho{\rm -str})}$ feeds the first charged harmonic through the $\sigma^z$ expansion \eqref{eq:LTsigmaZ}.  The bosonized continuum ansatz for the actual charged weak signal can therefore be written as
\begin{align}
C_{+-}^{\rm cont}(L,T)
&=
A_{\rm kin}\,\partial_T G_{+-}^{(0)}(L,T)
+A_{\rm stag}\,(-1)^{n_B}G_{+-}^{(0)}(L,T)
+A_\rho\,G_{+-}^{(2)}(L,T)
+\cdots,
\label{eq:chargedContinuumCandidate}
\\
G_{+-}^{(0)}(L,T)
&\equiv
\expval{\mathcal O_{+1,0}(x_A,0)\,\mathcal O_{-1,0}(x_B,T)}_g,
\label{eq:Gpm0}
\\
G_{+-}^{(2)}(L,T)
&\equiv
\expval{\mathcal O_{+1,0}(x_A,0)\,\Bigl(\mathcal O_{-1,2}+\mathcal O_{-1,-2}\Bigr)(x_B,T)}_g,
\label{eq:Gpm2}
\end{align}
with nonuniversal matching coefficients $A_{\rm kin}$, $A_{\rm stag}$, and $A_\rho$.  Here $A_{\rm stag}$ denotes the effective continuum amplitude obtained by coarse-graining the lattice staggered-mass operator $m\,\epsilon_n Z_n$ onto the leading charged field.  It is therefore a matching coefficient in the charged weak signal channel, not a new microscopic mass parameter separate from the lattice staggered term.  The actual qubit channels are then the real and imaginary combinations of the charged pairings,
\begin{equation}
\eta_{\rm act,cont}^{(X)}=C_{+-}^{\rm cont}+C_{-+}^{\rm cont},
\qquad
\eta_{\rm act,cont}^{(Y)}=i\bigl(C_{+-}^{\rm cont}-C_{-+}^{\rm cont}\bigr),
\label{eq:actualChargedContinuumXY}
\end{equation}
with $C_{-+}^{\rm cont}$ obtained by charge conjugation.  In the concrete open-chain examples below, the separated signal appears predominantly in the $Y$ channel, which simply means that the imaginary opposite-charge pairing dominates over the real one.

This charged-sector identification changes the expected gapped scale of the actual minimal protocol.  Because $\mathcal O_{\pm1,0}$ carry unit topological charge, their bulk form-factor expansion is governed by one-soliton intermediate states rather than by the neutral current channel.  At short distance the critical power is fixed by Eq.~\eqref{eq:chargedUVScaling}.

The $\eta_{\rm XXZ}$-dependent power in Eq.~\eqref{eq:chargedUVScaling} controls the critical ultraviolet scaling of the charged field.  The $r^{-1/2}e^{-Mr}$ behavior used below is the universal one-soliton infrared tail of a massive charged propagator.  The comparison in this subsection therefore addresses the large distance structure of the charged sector rather than the full crossover from the critical ultraviolet regime to the massive infrared regime.

The charged-sector comparison is restricted to the universal massive one-particle infrared tail,
\begin{equation}
K_0(Mr)\sim \sqrt{\frac{\pi}{2Mr}}\,e^{-Mr},
\qquad
K_1(Mr)\sim \sqrt{\frac{\pi}{2Mr}}\,e^{-Mr},
\qquad Mr\gg1,
\label{eq:besselIRCharged}
\end{equation}
so the fixed $r^{-1/2}$ power in the boundary state comparison below is the infrared one-soliton tail.  Accordingly,
\begin{equation}
G_{+-}^{(0)}(r)\sim r^{-1/2}e^{-Mr},
\qquad
\eta_{\rm act}^{(X/Y)}(r)\sim r^{-1/2}e^{-Mr},
\qquad r\to\infty,
\label{eq:chargedAsymptoticCandidate}
\end{equation}
with $M$ the soliton mass.  The boundary amplitudes and the full crossover from the critical $\eta_{\rm XXZ}$-dependent power to the massive infrared tail are not determined here.  The charged sector identified above is therefore the part that must be computed in any fuller continuum treatment of the minimal qubit protocol.

\begin{figure}[t]
  \centering
  \safeincludegraphics[width=0.98\linewidth]{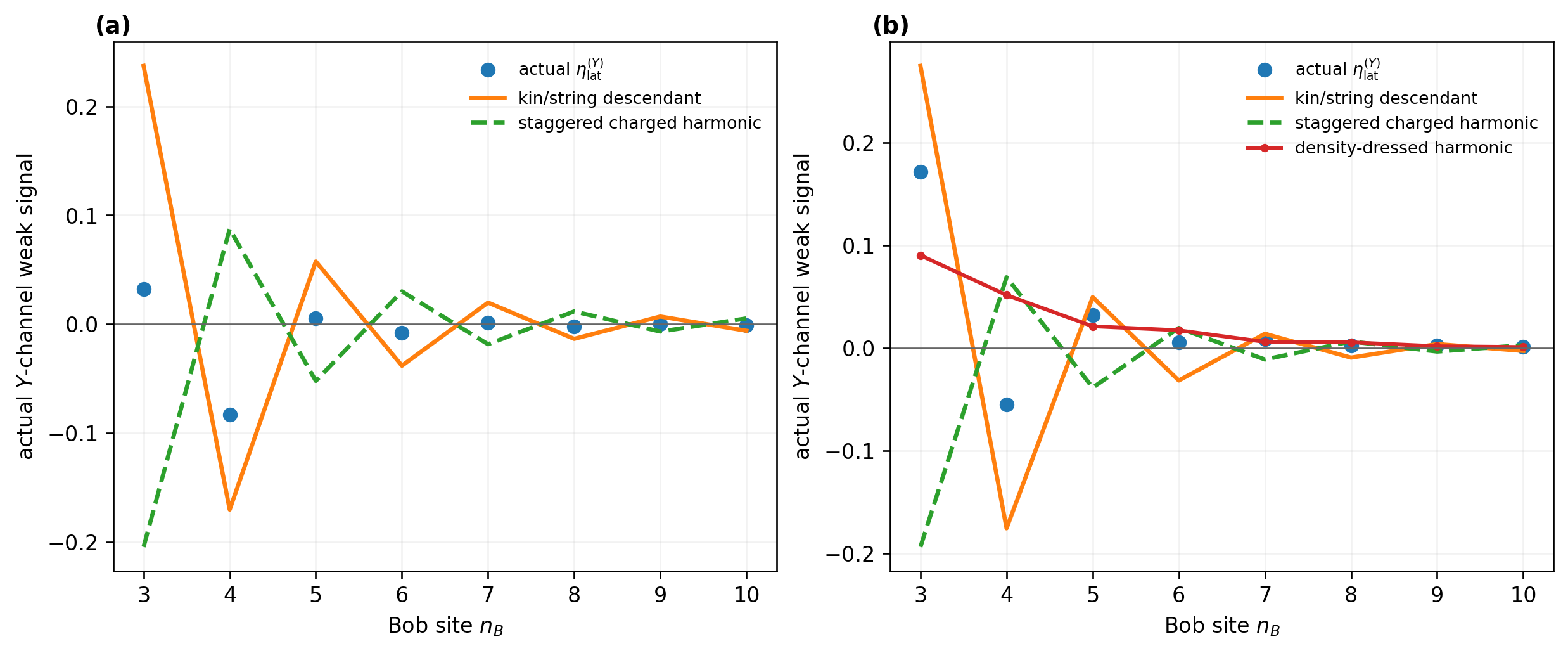}
  \caption{Charged-sector anatomy of the actual weak signal.  (a) free massive open chain with $\Delta(g)=0$.  The separated Bob-side $Y$ signal is produced by the interference between the leading charged descendant (the continuum image of the kinetic/string term) and the staggered charged harmonic.  (b) interacting open chain with $\Delta(g)=0.3$.  The same two contributions remain, and the interaction adds the density-dressed harmonic associated with the first charged harmonic generated by $\sigma^z$ dressing.  In both panels $N=10$, $m=0.4$, and Alice is fixed at $n_A=2$.  These data provide the lattice evidence underlying Eqs.~\eqref{eq:chargedContinuumCandidate}--\eqref{eq:actualChargedContinuumXY}.}
  \label{fig:ikedaChargedSector}
\end{figure}

Fig.~\ref{fig:ikedaChargedSector} confirms the operator content numerically.  On a free massive open chain ($\Delta(g)=0$), the separated actual $Y$ signal is generated by the interference of the leading charged descendant and the staggered charged harmonic, with no density-dressed piece.  Turning on $\Delta(g)$ leaves the same two pieces in place and adds the density-dressed harmonic predicted by \eqref{eq:etaLatYdecomp} and \eqref{eq:LTsigmaZ}.  The exact opposite-charge reconstruction \eqref{eq:etaXYfromChargedPairs} is also satisfied numerically to machine precision in the interacting example used for Fig.~\ref{fig:ikedaSelectionRule}.

\subsubsection{\texorpdfstring{Boundary comparison for $\eta_Y$.}{Boundary comparison for etaY.}}
The operator identities above determine the sector content of the actual nonzero weak signal, but not the interacting boundary correlator of the charged semi-local fields.  The comparison in this subsection uses a minimal boundary state description in which the boundary sign is fixed by the reflection properties of the charged sector, while the nonuniversal bulk amplitudes are obtained from an interior calibration.

In integrable boundary QFT the boundary state is built from opposite rapidities, and the boundary state amplitude is obtained from the reflection factor by the crossed relation $K(\theta)=R(i\pi/2-\theta)$ \cite{GhoshalZamolodchikov1994}.  For the SG charged sector the classical and semiclassical half-line analysis shows that Dirichlet boundaries reflect a kink as a kink, whereas Neumann boundaries reflect a kink as an anti-kink \cite{SaleurSkorikWarner1995}.  In the conformal / massless limit the $\phi_0=0$ Dirichlet problem also reduces to the odd-extension image law \cite{Cardy1984,SaleurSkorikWarner1995}.  Using the boundary reflection relation and exact boundary exponential expectation values \cite{FateevLukyanovZamolodchikovZamolodchikov1997}, one finds the minimal one-soliton charged image sign
\begin{equation}
 s_{\rm D}=+1,
 \qquad
 s_{\rm N}=-1.
 \label{eq:chargedReflectionSigns}
\end{equation}
Only this sign is taken from the boundary data, and it is not fitted to the edge data.  The charged comparison then keeps only the universal one-soliton infrared propagator on the half line and uses the minimal Euclidean boundary-state kernel
\begin{equation}
\mathcal G_{\chi,s}(x_A,x_B)
\equiv
K_0(M_\chi L)
+
 s\,K_0(M_\chi S),
\qquad
L\equiv x_B-x_A,
\qquad
S\equiv x_A+x_B,
\label{eq:chargedBoundaryStateKernel}
\end{equation}
where $K_\nu$ denotes the modified Bessel function of the second kind.  Its derivative, needed for the charged descendant piece, is
\begin{equation}
\partial_L\mathcal G_{\chi,s}(x_A,x_B)
=
-M_\chi K_1(M_\chi L)
-
s\,M_\chi K_1(M_\chi S).
\label{eq:chargedBoundaryStateKernelDeriv}
\end{equation}
At large distance this kernel reproduces Eq.~\eqref{eq:besselIRCharged}, so the fixed $r^{-1/2}$ power used here is the universal massive infrared tail.  The $\eta_{\rm XXZ}$ dependence remains in the operator identification and in the nonuniversal matching coefficients, not in the asymptotic Bessel exponent itself.  Because the actual separated $Y$ signal is already organized in Eq.~\eqref{eq:etaLatYdecomp} as a leading charged descendant plus staggered and density-dressed harmonics, the continuum ansatz used in the comparison is
\begin{equation}
\eta^{\rm cont}_{Y,s}(n_A,n_B)
=
B_{\rm kin}^{\rm bulk}\,\partial_L\mathcal G_{\chi,s}(x_A,x_B)
+
(-1)^{n_B}B_{\rm stag}^{\rm bulk}\,\mathcal G_{\chi,s}(x_A,x_B)
+
B_{\rho}^{\rm bulk}\,\mathcal G_{\chi,s}(x_A,x_B),
\label{eq:etaYBoundaryStatePrediction}
\end{equation}
with $B_{\rho}^{\rm bulk}=0$ when $\Delta(g)=0$.  The nonuniversal constants in Eq.~\eqref{eq:etaYBoundaryStatePrediction} are obtained from an interior calibration rather than from an edge fit.  We determine $M_\chi^{\rm bulk}$, $B_{\rm kin}^{\rm bulk}$, $B_{\rm stag}^{\rm bulk}$, and $B_{\rho}^{\rm bulk}$ once from an interior geometry ($n_A=3$) using the same bulk charged tail but with the image term removed.  In the free case this means three bulk parameters $(M_\chi^{\rm bulk},B_{\rm kin}^{\rm bulk},B_{\rm stag}^{\rm bulk})$ calibrated from $10$ separated interior points, while in the interacting case it means four bulk parameters $(M_\chi^{\rm bulk},B_{\rm kin}^{\rm bulk},B_{\rm stag}^{\rm bulk},B_{\rho}^{\rm bulk})$ calibrated from $8$ such points.  Those bulk-calibrated coefficients are then transferred unchanged to the edge geometry $n_A=1$.  The resulting comparison uses no fitted boundary amplitude.  The only boundary input is the sign $s=\pm1$ inherited from the Dirichlet/Neumann charged reflection rule.

\begin{figure}[t]
  \centering
  \safeincludegraphics[width=0.98\linewidth]{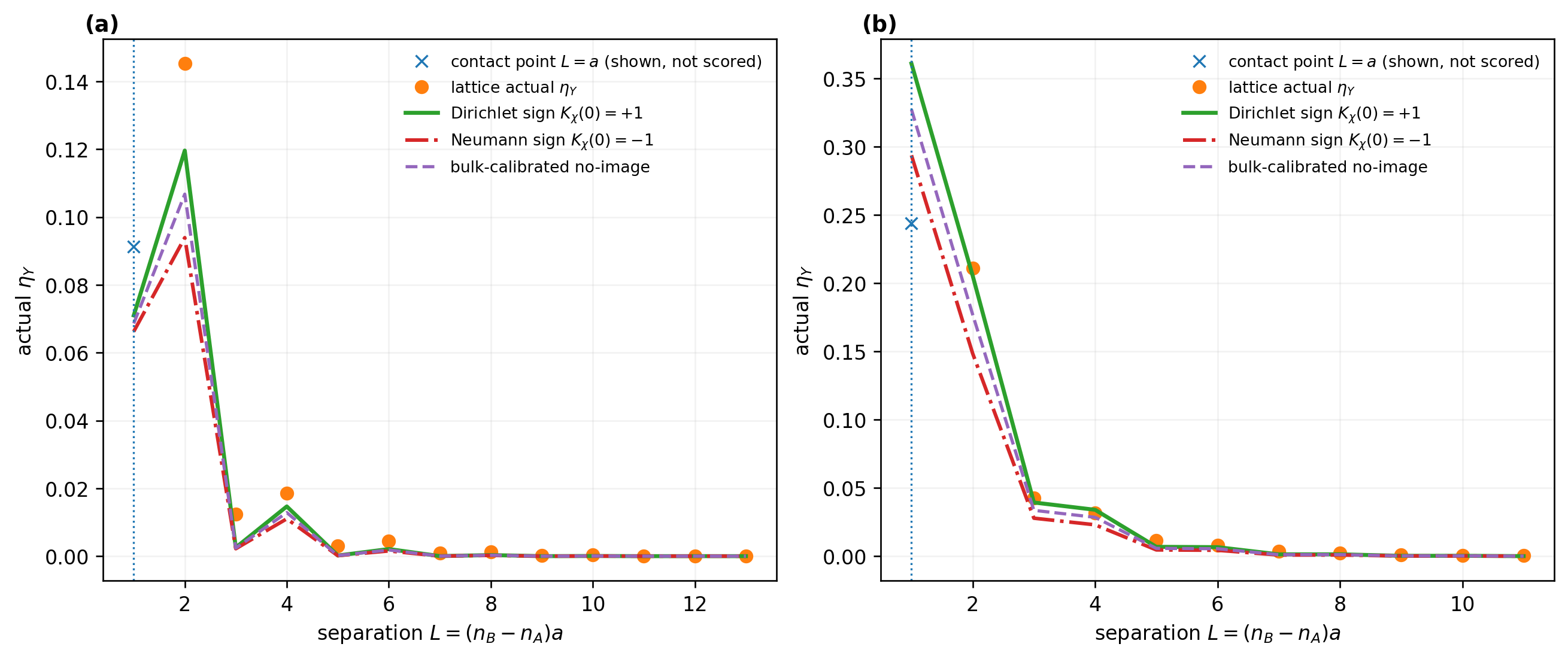}
  \caption{Boundary comparison for the charged $\eta_Y$ signal using the sign choice of Eqs.~\eqref{eq:chargedReflectionSigns}--\eqref{eq:etaYBoundaryStatePrediction}.  The charged bulk coefficients are calibrated once in the interior geometry $n_A=3$ with the image term removed and then transferred unchanged to the edge geometry $n_A=1$.  No boundary amplitude is fitted at the edge.  The solid curve uses the Dirichlet sign $K_\chi(0)=+1$, the dash-dot curve the Neumann sign $K_\chi(0)=-1$, and the dashed curve the bulk-only result without an image term.  The point $L=a$ is shown for completeness but excluded from the separated-window RMS comparison.  (a) free massive open chain ($N=14$, $m=0.4$, $\Delta(g)=0$).  (b) interacting open chain ($N=12$, $m=0.4$, $\Delta(g)=0.3$).  In both cases the Dirichlet sign improves the separated $\eta_Y$ profile, while the Neumann sign worsens it.}
  \label{fig:chargedVertexBridge}
\end{figure}

Fig.~\ref{fig:chargedVertexBridge} shows the result. In the free massive example ($N=14$, $m=0.4$, $\Delta(g)=0$), the bulk parameters are calibrated on $10$ separated interior points and then transferred unchanged to the edge geometry.  In the interacting example ($N=12$, $m=0.4$, $\Delta(g)=0.3$), the same procedure uses $8$ interior points.  In both cases the Dirichlet sign improves the separated-window agreement relative to the bulk-only curve, whereas the Neumann sign worsens it.  The comparison is qualitative but nontrivial.  Once the bulk amplitudes are fixed away from the edge, the edge profile favors the charge-preserving Dirichlet sign.

This comparison does not include the exact boundary form factors of $\mathcal O_{\pm1,0}$ and $\mathcal O_{\pm1,\pm2}$ or the corresponding ultraviolet and infrared normalization of their boundary amplitudes.

\begin{figure}[t]
  \centering
  \safeincludegraphics[width=0.82\linewidth]{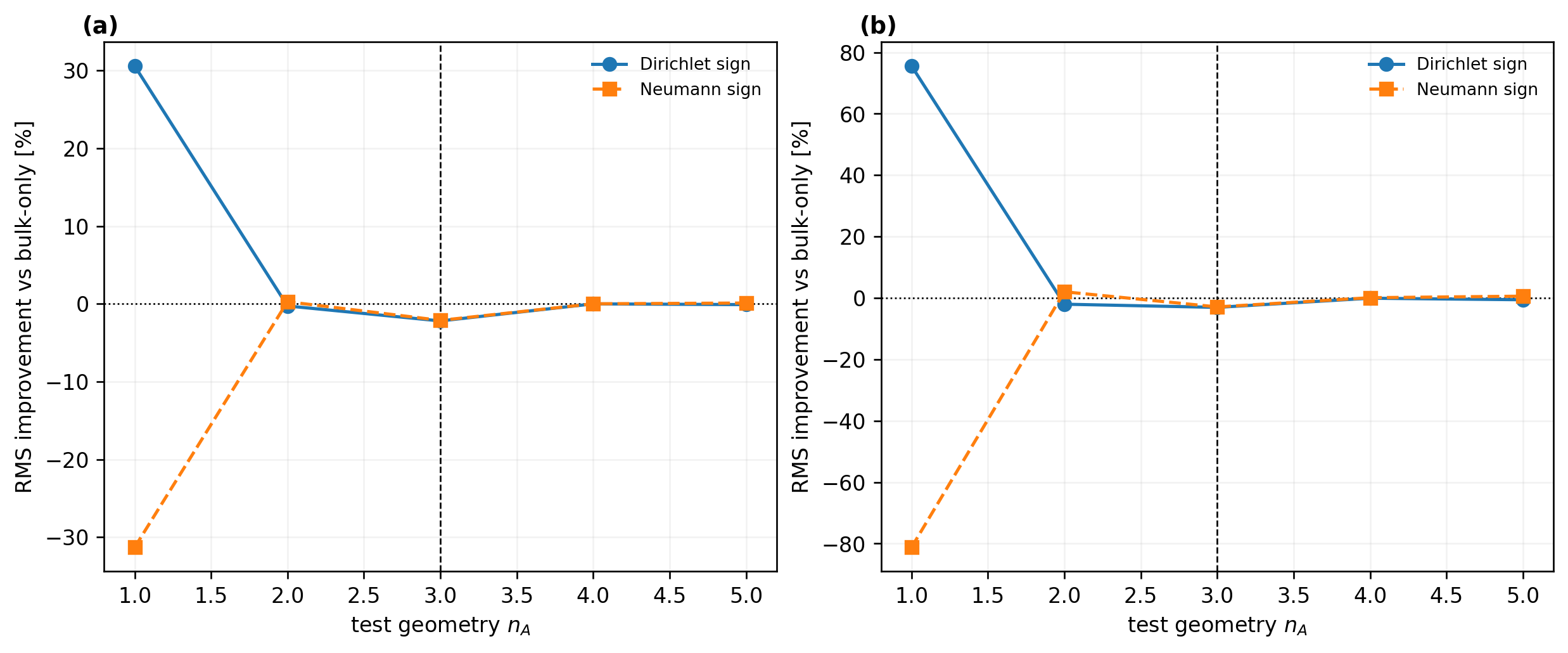}
  \caption{Edge sensitivity of the charged boundary comparison.  Panel~(a) shows the free chain and panel~(b) the interacting chain.  The bulk calibration is fixed at $n_A=3$, and the test geometry is moved from the edge inward through the scan $n_A=1,\ldots,5$.  The vertical axis shows the percentage RMS change relative to the bulk-only curve.  The sign discrimination is large at the edge and quickly collapses away from it.  The edge/off-edge separation is roughly two orders of magnitude in the free case and about forty in the interacting case.}
  \label{fig:chargedBoundaryDiagnostics}
\end{figure}

Keeping the same interior calibration fixed at $n_A=3$ and moving the test geometry inward makes the spatial range of the comparison explicit.  In the scan $n_A=1,\ldots,5$, the sign discrimination is strongest at the edge.  At $n_A=1$ the Dirichlet choice improves the RMS relative to the bulk-only curve by about one third in the free example and by about three quarters in the interacting example, while the Neumann choice worsens it by a comparable amount.  One site inward the changes are already near zero in the free case and only at the few-percent level in the interacting case, and the remaining inward points show similarly weak discrimination.  The edge/off-edge sign-separation measure is therefore larger at the boundary by roughly two orders of magnitude in the free example and by roughly forty times in the interacting example.  Fig.~\ref{fig:chargedBoundaryDiagnostics} shows that the sign discrimination is localized near the edge in the short chains studied here.

\subsubsection{Implications near criticality.}
Ref.~\cite{Ikeda2023} observed that the extracted energy peaks near phase-transition lines.  The continuum analysis of the neutral current sector on the same lattice Hamiltonian gives a closely related statement.  Within the neutral current protocol of Eqs.~\eqref{eq:latticeCurrentPOVM}--\eqref{eq:etaJZequalsJJ}, the weak signal obeys the same crossover law as the continuum coefficient $\eta_1$:
\begin{equation}
\eta_{JJ,\ell}^{\rm lat}(L_{\rm lat},T)
\sim
\begin{cases}
L_{\rm lat}^{-2}, & L_{\rm lat}\ll \xi_{\rm corr}^{(j)} \quad \text{(critical or crossover window)},\\[4pt]
L_{\rm lat}^{-\gamma_j}\exp\!\left(-\dfrac{L_{\rm lat}}{\xi_{\rm corr}^{(j)}}\right), & L_{\rm lat}\gg \xi_{\rm corr}^{(j)} \quad \text{(gapped bulk asymptotic regime)}.
\end{cases}
\label{eq:latticeCurrentProxyCrossover}
\end{equation}
Consequently the corresponding weak signal energy scales as
\begin{equation}
E_{B,{\rm curr}}^{(J)}(L_{\rm lat})
\propto
\begin{cases}
L_{\rm lat}^{-4}, & L_{\rm lat}\ll \xi_{\rm corr}^{(j)},\\[4pt]
L_{\rm lat}^{-2\gamma_j}\exp\!\left(-\dfrac{2L_{\rm lat}}{\xi_{\rm corr}^{(j)}}\right), & L_{\rm lat}\gg \xi_{\rm corr}^{(j)}.
\end{cases}
\label{eq:latticeCurrentEnergyProxy}
\end{equation}
These are bulk crossover laws for the neutral current sector.  The boundary-to-bulk geometry of Fig.~\ref{fig:latticeCurrentBridge} serves only to select the free-point continuum reference kernel and is separate from the asymptotic statement above.  Increasing $\xi_{\rm corr}^{(j)}$ near a critical line therefore enhances the neutral current QET mechanism available on the same lattice Hamiltonian.  This does not by itself explain the optimized qubit peak of the original Alice-$X$ protocol, because Eq.~\eqref{eq:etaLatZvanishes} shows that the neutral current sector does not contribute there.  The nonzero signal of the original protocol instead comes from the charged string/disorder and staggered/density sectors in the $X/Y$ channels.  The charged asymptotic form \eqref{eq:chargedAsymptoticCandidate} suggests a one-soliton bulk scale $e^{-Mr}$ for that signal, whereas the neutral current sector on the repulsive side is governed by $e^{-2Mr}$.  A complete continuum account of the original protocol therefore requires the interacting charged boundary correlators and their amplitudes.
\section{Discussion}
\label{sec:discussion}
The continuum analysis yields an exact quadratic functional in the current sector, with $\tilde D_A^{\,2}=\mathbf 1$ and a leading coefficient $\eta_1$ determined by a current correlator.  In the gapped theory, the spectral and form-factor representation of $\eta_1$ is exact at leading order in the weak measurement, while the asymptotic formulas apply in the regime $r/\xi_{\rm corr}^{(j)}\gg 1$.

For the single-site Pauli protocol of Ref.~\cite{Ikeda2023}, the lattice analysis gives a different sector structure.  Under the stated finite chain assumptions, the Bob-side neutral current channel vanishes for Alice's $X_{n_A}$ measurement, the separated $X$ channel is suppressed in the real basis convention, and the surviving signal is organized by opposite-charge pairings.  The nonzero weak signal of the original protocol is therefore not described by the neutral current sector.

The direct continuum--lattice comparison instead concerns the neutral current protocol defined on the same lattice Hamiltonian.  Its weak signal reduces exactly to a coarse-grained current correlator, and its leading extracted energy has the same quadratic QET structure with an explicit local energetic cost.  At the free open-chain point, the boundary comparison favors the Dirichlet image kernel as the natural continuum reference for this sector.

For the original qubit protocol, the missing continuum ingredient is the charged correlator with open boundaries.  The lattice sector analysis identifies the relevant charged scaling fields and shows that the Dirichlet sign gives the better edge comparison in the short chains studied here.  A full continuum treatment still requires the interacting boundary correlators and the ultraviolet and infrared normalization of the charged operators.
\section{Conclusions}
\label{sec:conclusions}
We studied QET in the bosonized massive Thirring model from both continuum and lattice perspectives.  In canonical normalization, the trigonometric POVM of Refs.~\cite{Hotta2008,HottaEM2010} satisfies $\widetilde D_A^{\,2}=\mathbf 1$, so the leading weak measurement signal is governed by a smeared current correlator.  This gives an explicit continuum protocol in the current sector, with the gapless law $r^{-4}$ for the extracted energy and the gapped large-distance behavior $E_B^{\max}\sim r^{-2\gamma_j}e^{-2M_{\rm eff}r}$.

We then analyzed the open-chain single-site Pauli protocol of Ref.~\cite{Ikeda2023}.  The central lattice result is that the Bob-side neutral current channel vanishes exactly for Alice's $X_{n_A}$ measurement by a $U(1)$ selection rule.  The separated signal therefore lies in charged $X/Y$ sectors, which explains why the original protocol does not couple directly to the continuum current sector.

A neutral current protocol on the same lattice Hamiltonian gives the corresponding lattice version of that continuum sector.  Its weak signal is exactly a coarse-grained current correlator, and its leading extracted energy follows the expected quadratic QET law.  Related chirality-based projective protocols of Ref.~\cite{IkedaBeyond2025} can activate local current and charge on the lattice, but the protocol studied here isolates the neutral conserved current channel and matches it to the continuum weak binary POVM.

For the charged sector of the original protocol, the relevant continuum operators are $\mathcal O_{\pm1,0}$ and $\mathcal O_{\pm1,\pm2}$.  The short-chain edge comparison favors the Dirichlet sign in the boundary state description.  Extending this part of the analysis requires the corresponding interacting charged correlators with open boundaries and their ultraviolet and infrared normalization.

\appendix
\section{Relation to the Lukyanov--Zamolodchikov normalization}
\label{app:LZnormalization}
Some SG references use the field $\varphi$ normalized by
\begin{equation}
S=\int d^2x\,\left[\frac{1}{16\pi}(\partial\varphi)^2-2\mu\bigl(1-\cos(\beta\varphi)\bigr)\right].
\end{equation}
The relation to the canonical field used in the main text is
\begin{equation}
\varphi=\sqrt{8\pi}\,\phi,
\qquad
\beta_C^2=8\pi\beta^2.
\end{equation}
In this convention the current map becomes
\begin{equation}
 j^\mu=-\frac{\beta}{2\pi}\epsilon^{\mu\nu}\partial_\nu\varphi,
\end{equation}
which is equivalent to \eqref{eq:currentmapCanonical}.  The BKT line $\beta_C^2=8\pi$ is therefore the same as $\beta^2=1$.

\bibliographystyle{utphys}
\bibliography{references_verified_v5}

\end{document}